%% file: main.tex
\documentclass[sigconf]{acmart}

\usepackage{fancyhdr}
\usepackage{url}
\usepackage{microtype}

\usepackage{algorithmic}
\usepackage{textcomp}
\usepackage{xcolor}
\usepackage{enumitem}
\usepackage{tikz}
\usepackage{siunitx}
\usepackage{soul}
\sethlcolor{yellow}
\usepackage{hyperref}

\usepackage{def}

\AtBeginDocument{%
  \providecommand\BibTeX{{%
    \normalfont B\kern-0.5em{\scshape i\kern-0.25em b}\kern-0.8em\TeX}}}

\copyrightyear{2021}
\acmYear{2021}
\setcopyright{acmlicensed}\acmConference[MICRO '21]{MICRO-54: 54th Annual IEEE/ACM International Symposium on Microarchitecture}{October 18--22, 2021}{Virtual Event, Greece}
\acmBooktitle{MICRO-54: 54th Annual IEEE/ACM International Symposium on Microarchitecture (MICRO '21), October 18--22, 2021, Virtual Event, Greece}
\acmPrice{15.00}
\acmDOI{10.1145/3466752.3480109}
\acmISBN{978-1-4503-8557-2/21/10}

\begin{document}

\title[IceClave: A Trusted Execution Environment for In-Storage Computing]{IceClave: A Trusted Execution Environment\\ for In-Storage Computing}

\input{authors}

\input{abstract}



\maketitle

\input{introduction}
\input{background}

\input{threatmodel}
\input{design}

\input{implt}

\input{evaluation}
\input{relatedwork}
\input{conclusion}

\input{ack}

\bibliographystyle{ACM-Reference-Format}
\bibliography{references,ref}

\end{document}

%% file: authors.tex






\author{Luyi Kang}
\authornote{Co-primary authors.}
\authornote{Work done while visiting the Systems Platform Research Group at UIUC.}
\affiliation{%
  \institution{University of Maryland, College Park}
  \streetaddress{Address}
  \city{}
  \state{}
  \country{}}

\author{Yuqi Xue}
\authornotemark[1]
\affiliation{%
  \institution{UIUC}
  \streetaddress{Address}
  \city{}
  \state{}
  \country{}}

\author{Weiwei Jia}
\authornotemark[1]
\affiliation{%
  \institution{UIUC}
  \streetaddress{Address}
  \city{}
  \state{}
  \country{}}

\author{Xiaohao Wang}
\authornote{Now at NVIDIA.}
\affiliation{%
  \institution{UIUC}
  \streetaddress{Address}
  \city{}
  \state{}
  \country{}}

\author{Jongryool Kim}
\affiliation{%
  \institution{SK Hynix}
  \streetaddress{Address}
  \city{}
  \state{}
  \country{}}

\author{Changhwan Youn}
\affiliation{%
  \institution{SK Hynix}
  \streetaddress{Address}
  \city{}
  \state{}
  \country{}}

\author{Myeong Joon Kang}
\affiliation{%
  \institution{SK Hynix}
  \streetaddress{Address}
  \city{}
  \state{}
  \country{}}

\author{Hyung Jin Lim}
\affiliation{%
  \institution{SK Hynix}
  \streetaddress{Address}
  \city{}
  \state{}
  \country{}}

\author{Bruce Jacob}
\affiliation{%
  \institution{University of Maryland, College Park}
  \streetaddress{Address}
  \city{}
  \state{}
  \country{}}

\author{Jian Huang}
\affiliation{%
  \institution{UIUC}
  \streetaddress{Address}
  \city{}
  \state{}
  \country{}}

\renewcommand{\shortauthors}{L. Kang, Y. Xue, W. Jia, X. Wang, J. Kim, C. Youn, M. Kang, H. Lim, B. Jacob, J. Huang}

%% file: abstract.tex
\begin{abstract}
In-storage computing with modern solid-state drives (SSDs)   
enables developers to offload programs from the host to the 
SSD. It has been proven to be an effective approach to alleviate the I/O bottleneck. 
To facilitate in-storage computing, many frameworks have been proposed. 
However, few of them treat the 
in-storage security as the first citizen. 
Specifically, since modern SSD controllers do not have a trusted execution environment, 
an offloaded (malicious) program could steal, modify, and even destroy the data stored in 
the SSD. 

In this paper, we first investigate the attacks that could be conducted by offloaded in-storage 
programs. To defend against these attacks, we build a lightweight trusted
execution environment, named IceClave for in-storage computing. 
IceClave enables security isolation between in-storage programs and flash management functions that 
include flash address translation, data access control, and garbage collection, with TrustZone extensions.  
IceClave also achieves security isolation between in-storage programs by enforcing  
memory integrity verification of in-storage DRAM with low overhead. To protect data loaded from flash chips,   
IceClave develops a lightweight data encryption/decryption mechanism in flash controllers.  
We develop IceClave with a full system simulator. 
We evaluate IceClave with a variety of 
data-intensive applications such as databases.
Compared to state-of-the-art in-storage computing approaches, 
IceClave introduces only 7.6\% performance overhead, while enforcing security isolation in 
	the SSD controller with minimal hardware cost.  
IceClave still keeps the performance benefit of in-storage computing by delivering up to 2.31$\times$
better performance than the conventional host-based trusted computing approach.

\end{abstract}

%% file: introduction.tex
\section{Introduction}
\label{sec:intro}
In-storage computing has been a promising technique for accelerating data-intensive 
applications, especially for large-scale data processing and analytics~\cite{summarizer, willow, neardata, biscuit, deepstore:micro2019, graphssd:isca2019, GraFBoost, 
Do:2013:QPS:2463676.2465295, Tiwari:2013:AFT:2591272.2591286, polardb}. 
It moves computation closer to the data stored in the storage devices like  
flash-based solid-state drives (SSDs), such that it can overcome the  
I/O bottleneck by significantly reducing the amount of data transferred between 
the host machine and storage devices. 
As modern SSDs are employing multiple general-purpose embedded processors 
and large DRAM in their controllers, it becomes feasible to enable in-storage 
computing in reality today. 

To facilitate the wide adoption of in-storage computing, a variety of frameworks have 
been proposed. 
For instance, Willow~\cite{willow} enabled developers to offload 
code from the host machine to the SSD via RPC protocols, and Biscuit~\cite{biscuit} 
developed an in-storage runtime system for supporting multiple in-storage computing 
tasks following the MapReduce computing model. 
All these prior works show the great potential of in-storage 
computing for accelerating data processing in data centers. 
However, most of them~\cite{summarizer,willow,Do:2013:QPS:2463676.2465295,smartssd, ActiveFlash, graphssd:isca2019,Tiwari:2012:RDM:2387869.2387873,deepstore:micro2019} focus on the performance and programmability, few of them treat the security as the first 
citizen in their design and implementation, which imposes great threat to the user data and 
SSD devices, and further hinders its widespread adoption. 

As in-storage processors operate independently from the host machine, and modern  
SSD controller does not provide a trusted execution environment (TEE) for programs running inside the SSD,  
they pose severe security threats to user data and flash chips. 
To be specific, a piece of offloaded (malicious) code
could (1) manipulate the mapping table in the flash translation layer (FTL) to mangle  
the data management of flash chips, (2) access and destroy data belonging 
to other applications, and (3) steal and modify the memory of co-located 
in-storage programs at runtime.

To overcome these security challenges, 
as developed in state-of-the-art in-storage computing frameworks~\cite{biscuit, willow}, 
we can simplify the runtime system by maintaining 
a copy of the privilege information in the DRAM of SSD controllers (SSD DRAM) and 
enforcing permission checks for in-storage programs. 
However, such a solution still suffers from many security vulnerabilities.  
For example, 
a malicious offloaded 
program can exploit memory vulnerabilities such as buffer overflow~\cite{sok, pointerguard:security2003, bufferoverflow} 
to enable privilege escalation to access and modify the cached mapping table of FTL in the SSD DRAM;  
adversaries can steal and modify intermediate data and results generated by the co-located 
in-storage programs via physical attacks such as cold-boot attack, bus snooping attack, 
and replay attack~\cite{physicalattack:blackhat2014, morphable:micro18, ssdflaw}.  
An alternative approach is to adopt Intel's Software Guard Extension (SGX) as a drop-in solution. Unfortunately, modern in-storage 
processor architectures do not support SGX techniques. And the SGX approach still suffers 
from significant performance overhead~\cite{shieldstore:eurosys2019,vault:asplos18,www:intelsgx,vc3:oakland2015,enclavedb,scone:osdi2016}, 
which cannot be afforded by in-storage computing and SSD controllers today.

\begin{figure}[t]
\centering
\includegraphics[width=0.88\linewidth]{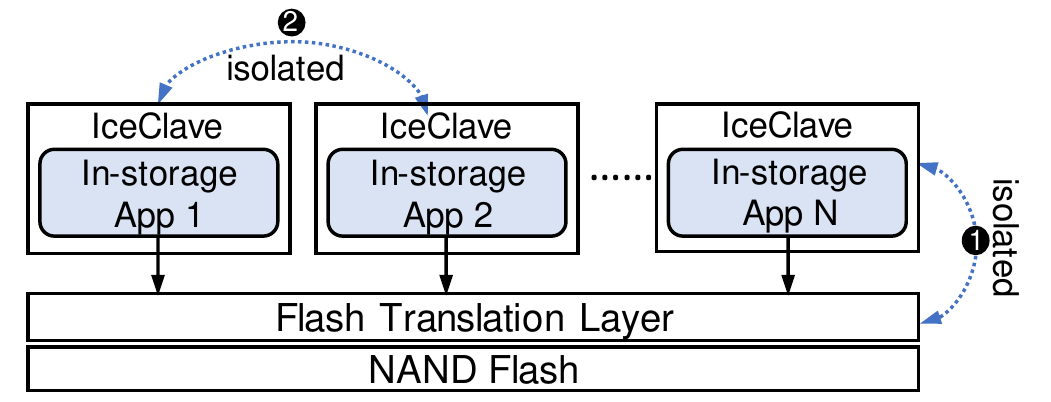}
\vspace{-1.5ex}
\caption{IceClave enables in-storage TEEs to achieve security isolation between in-storage programs, 
	FTL, and flash chips. The shaded components are untrusted.}
\label{fig:goals}
\vspace{-3ex}
\end{figure}

Therefore, providing a secure, lightweight, and trusted execution environment  for in-storage computing 
is an essential step towards its widespread adoption. 
Ideally, we wish to enjoy the performance benefits of in-storage computing, while 
enforcing the security isolation between in-storage programs, the core FTL functions, and physical 
flash chips, as demonstrated in Figure~\ref{fig:goals}.  

To this end, we present IceClave, a trusted execution environment for in-storage computing. 
Unlike generic TEE solutions such as SGX~\cite{www:intelsgx}, IceClave is designed specifically for modern SSD controllers 
and in-storage programs, with considering the unique flash properties and in-storage workload characteristic. 
With ensuring the security isolation, IceClave includes (1) a new memory protection scheme to reduce the context switch overhead 
incurred by flash address translations; (2) an optimization technique for securing in-storage DRAM for in-storage programs by taking 
advantage of the fact that most in-storage applications are read intensive; (3) a stream 
cipher engine for securing data transfers between storage processors and flash chips, with low performance overhead and energy consumption; 
and (4) a runtime system for managing the life cycle of in-storage TEEs.

Specifically, to achieve the security isolation between the in-storage 
program and the FTL (\circleb{1} in Figure~\ref{fig:goals}), 
we extend the TrustZone of ARM processors available in a majority of SSD controllers.  
IceClave executes core FTL functions such as garbage collection and wear leveling in the secure world, 
and runs the offloaded programs in the normal world,  
such that offloaded programs cannot intervene the flash management. 
To protect the mapping table of FTL with low overhead,  
we introduce a protected memory region in the normal world, and place the address mapping table in it to avoid context switches 
for flash address translation. Therefore, only protected FTL functions can update the mapping table, and offloaded in-storage programs 
can only read it for address translation with enforced permission checks.

In order to achieve the security isolation between in-storage programs 
(\circleb{2} in Figure~\ref{fig:goals}), 
we build in-storage TEEs to host the offloaded programs, and enforce  
data encryption and memory integrity checks in both data communication and 
processing for in-storage programs. 
Since most in-storage computing workloads are read intensive, 
IceClave mainly needs to conduct the integrity check for the intermediate 
data and results generated by in-storage programs, which does not introduce much 
performance overhead to the in-storage program execution.  
To secure flash pages read by in-storage programs running in the in-storage TEE,  
we also integrate a lightweight stream cipher engine into the SSD controller with minimum hardware cost. 
In summary, we make the following contributions in this paper:

\begin{itemize}[leftmargin=*]
\item We present a trusted execution environment for in-storage computing, in which 
	it protects core FTL functions from malicious in-storage programs with TrustZone extension.

\vspace{1ex}
\item We support security isolation between in-storage programs by enabling lightweight memory encryption 
and integrity verification for SSD DRAM. 
\vspace{1ex}
\item We show the required hardware and software extensions for IceClave are minimal and feasible to 
	be developed in modern SSD controllers. 
\vspace{1ex}
\item We develop a system prototype using an SSD FPGA board for a quantitative overhead study of IceClave, 
and a full system simulator for sensitivity analysis. 
\vspace{1ex}


\end{itemize}

Specifically, we implement IceClave with a full system simulator gem5~\cite{gem5}, and integrate an 
SSD simulator SimpleSSD~\cite{simplessd} and memory simulator USIMM~\cite{usimm} into it for supporting secure in-storage computing. 
We also develop a system prototype to verify the core functions of IceClave  
with a real-world OpenSSD Cosmos+ FPGA board~\cite{openssd-web}. 
We use a variety of data-intensive applications that include transactional databases to evaluate the efficiency  
of IceClave. 
Compared to state-of-the-art in-storage computing approaches,  
IceClave introduces only 7.6\% performance overhead to the in-storage runtime, 
while adding minimal area and energy overhead to the SSD controller. 
Our evaluation also demonstrates that 
IceClave can maintain the performance benefit of in-storage computing by delivering 
2.31$\times$ better performance on average than the conventional host-based computing approach.  


%% file: background.tex
\section{Background and Motivation}
\label{sec:motivation}
In this section, we first introduce the background of flash-based SSDs and 
in-storage computing, and discuss why it is desirable to have TEEs 
for in-storage computing. 

\begin{figure}[t]
    \centering
    \includegraphics[width=0.9\linewidth]{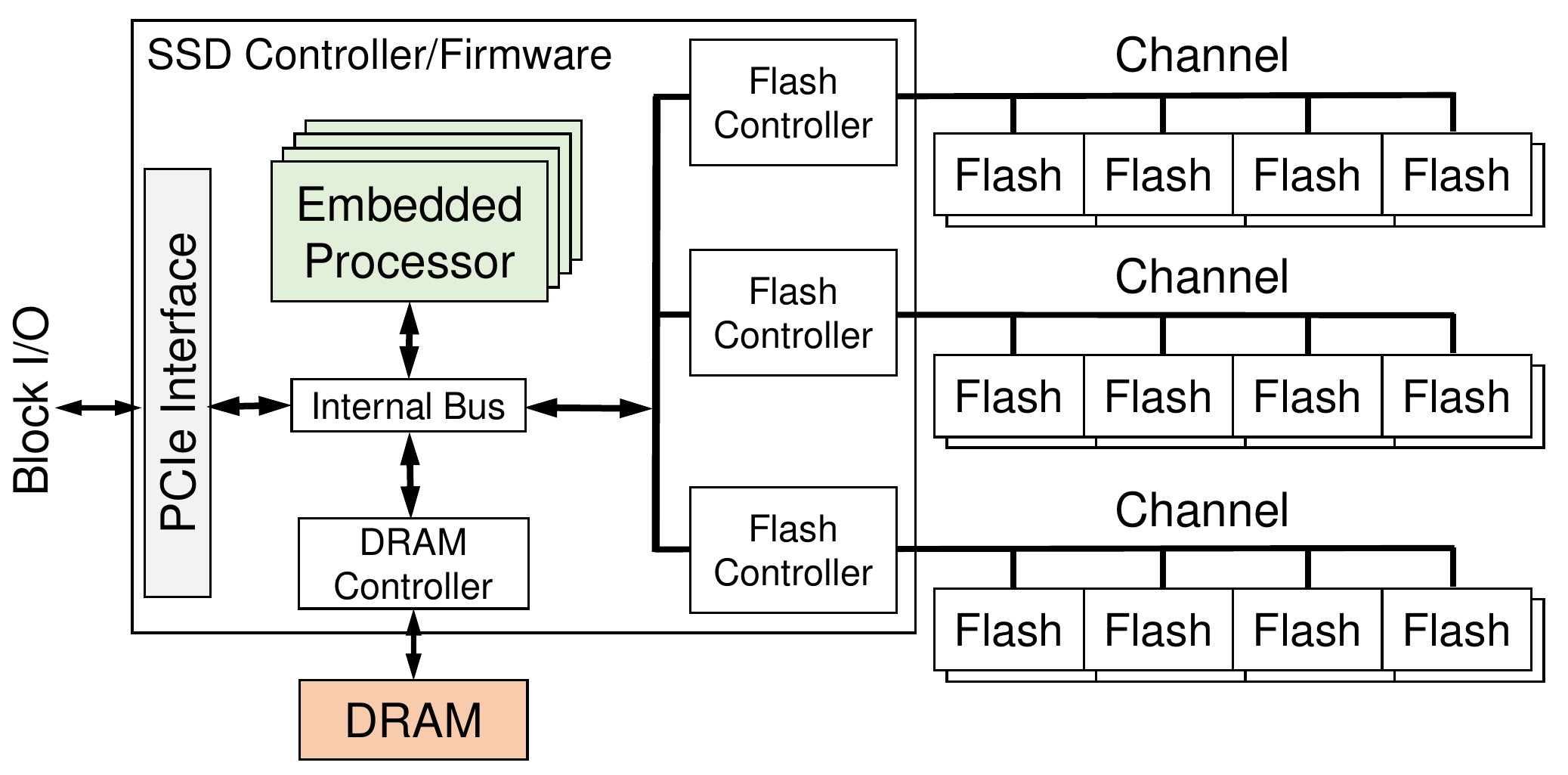}
    \vspace{-2ex}
    \caption{Internal architecture of flash-based SSDs.}
    \label{fig:ssd-arch}
    \vspace{-3ex}
\end{figure}

\subsection{Flash-Based Solid-State Drive}
\label{subsec:ssd}

The rapidly shrinking process technology has allowed SSDs to boost their performance
and capacity, which accelerates their adoption in commodity systems such as data 
centers today~\cite{flashblox, flashmap:isca2015, flatflash:asplos2019}.
We present a typical SSD architecture in Figure~\ref{fig:ssd-arch}.
An SSD has three major components: a set of flash memory packages,
an SSD controller having embedded processors with DRAM, and flash controllers~\cite{chen2011essential, 
deepstore:micro2019}. The flash packages are
organized in a hierarchical manner. Each SSD has multiple channels. 
Each channel is shared by multiple flash
packages. Each package consists of multiple flash chips. Each chip has multiple planes. Each plane
is divided into multiple flash blocks, each of which consists of multiple flash pages.

Due to the nature of flash memory,
when a free flash page is written once, that page is no longer available for future writes until
that page is erased.
However, erase operation can be performed only at a block 
granularity, which is time-consuming.
Thus, writes are issued to free pages that have been erased (i.e., out-of-place write) rather than
waiting for the expensive erase operation. 
Garbage collection (GC) will be performed later to clean the obsolete data in SSDs.
As each flash block has limited endurance,
it is important for blocks to age uniformly (i.e., wear leveling).
SSDs employ out-of-place write, GC, and wear leveling to overcome the shortcomings
of SSDs and maintain indirections for indexing the logical-to-physical address mapping.
All these are managed by the Flash Translation Layer (FTL) in the SSD controller.

\subsection{In-Storage Computing}
\label{subsec:instorage}
We have long recognized the inefficiency of traditional CPU-centric computing 
for data-intensive applications that need to transfer large amounts of 
data from storage.
The application performance is limited by the low bandwidth of 
PCIe interface between the host and SSD. 
To tackle this challenge, various in-storage computing 
approaches~\cite{biscuit, summarizer, willow, capi:2014:whitepaper, deepstore:micro2019} have been proposed.
With them, we can process data by exploiting SSD processors and high internal 
bandwidth of flash chips. 
Their significant performance benefits show that in-storage 
computing is a promising technique. 

However, as in-storage processors operate independently from the host,  
it poses security challenges to the adoption of in-storage computing, especially in 
the multi-tenancy setting where multiple application instances share the physical 
SSD~\cite{multissd:sigmod2014, documentdb:vldb2015, cloudstorage:osdi2012, libra:eurosys2014, 
slo:fast2015, heracles:isca2015}.

\subsection{In-Storage Vulnerabilities}
\label{subsec:why}
When specific code is offloaded to in-storage processors, a copy of the privilege information is transferred and maintained in the DRAM of the SSD controller~\cite{biscuit,willow}. 
Such a solution is developed under the assumption that the offloaded code has already 
known the address and size of the accessed data in advance.
However, 
adversaries can exploit in-storage software and firmware vulnerabilities such as 
buffer overflows~\cite{sok, pointerguard:security2003, bufferoverflow},  
and bus-snooping attack~\cite{ssdflaw} to achieve privilege escalation. After that, 
they can conduct various further attacks.  
We present them in the following. 

\begin{itemize}[leftmargin=*]
    \item {A malicious user can manipulate the intermediate data and output generated 
    	by in-storage programs via both software and physical attacks, causing incorrect computing results. }
\vspace{1ex}
	\item {A malicious program can 
		intercept FTL functions like GC and 
	        wear leveling in the SSD and mangle the flash management. This would cause data loss 
		or device destroyed. } 
\vspace{1ex}
	\item {A malicious user can steal user data stored in flash chips via physical attacks 
	like bus snooping attack, when 
		in-storage programs load data from flash chips to SSD DRAM. }
		\vspace{1ex}
		
\end{itemize}

To defend against these attacks, an alternative solution is to develop an OS or hypervisor for 
in-storage computing. However, due to the limited resources  
in the SSD controller, running a full-fledged OS can introduce significant overheads to the SSD and increase the attack surface, due to its large codebase.
Moreover, these techniques are not sufficient to defend against the aforementioned attacks such as 
board-level physical attacks.
As we move compute closer to storage devices, it is highly desirable to have a lightweight execution environment 
for this non-traditional computing paradigm.

Since SSDs were designed with the assumption they are hardware 
isolated from the host and purely used as storage rather than computing, 
modern computing systems do not provide secure runtime environment for in-storage computing.
As discussed in $\S$\ref{sec:intro}, a straightforward approach is to adopt the SGX-like solutions~\cite{www:intelsgx,keystone:eurosys2020}. 
However, this requires significant hardware changes and even replacement of storage processors available in modern SSDs. 
As SGX was developed as a generic framework for host machines, it is hard to achieve optimal performance for 
in-storage programs. 


%% file: threatmodel.tex
\section{Threat Model}
\label{sec:threatmodel}
In this work, we target the multitenancy where multiple application instances operate in the shared 
SSD. Following the threat models for cloud computing today
~\cite{heaven,komodo:sosp2017,sanctum:security2016,f1},
we assume the cloud computing platform has provided a secure channel for end users to offload 
their programs to the shared SSD. The related code-offloading techniques, such as secure RPC and libraries~\cite{biscuit, cstorage, willow}, have been deployed in cloud platforms~\cite{fpgacloud, amorphos, f1}. 
However, an offloaded program can include (hidden) malicious code. 

As for in-storage computing, 
we trust SSD vendors, 
who enable the execution of offloaded programs. 
We assume hardware vendors do not intentionally implant backdoor or malicious programs in their 
devices. However, as we deploy those computational SSDs in shared platforms (e.g., public cloud), 
we do not trust platform operators who could initiate board-level physical attacks such as 
bus-snooping and man-in-the-middle attacks, or exploit the host machine to steal or destroy data 
stored in SSDs. Similar to the threat model for SGX, 
we exclude software side-channel attacks such as cache timing, page table side-channel 
attacks~\cite{varys:atc2018}, and speculative attacks~\cite{kocher:sp2019} 
 from the threat model, since many of these attack 
approaches are cumbersome in reality~\cite{www:intelsgx}. 

We rely on the Error-Correction Code (ECC) available in flash controllers~\cite{tseng:dac2013, dftl} 
for ensuring the integrity of flash pages.
To defend against attacks from cloud computing platforms or malicious host OS, users are 
usually encouraged to encrypt their data before storing them in the SSD. 
However, their data 
would still be leaked during the in-storage computing procedure. Thus, we have a more conservative 
design for achieving the security goals of IceClave.

We believe our threat model is realistic.
First, as system-wide shared resource, SSDs have been widely used by multiple applications. Existing 
in-storage computing frameworks have enabled end users to offload their programs into the SSD. 
Second, once program is offloaded to the SSD, the in-storage program will escape the control 
of the host OS and initiate attacks in the new execution environment. Third, our threat model considers 
the potential physical attacks initiated by untrusted platform operators. 

To the best of 
our knowledge, this is the first TEE framework for in-storage computing.
{It aims to defend against three attacks: (1) the attack against co-located in-storage 
programs; (2) the attack against the core FTL functions; (3) the potential physical attack 
against the data loaded from flash chips and generated by in-storage programs. }

%% file: design.tex
\section{Design and Implementation}
\label{sec:design}
\begin{figure}
	\centering
	\includegraphics[width=0.8\linewidth]{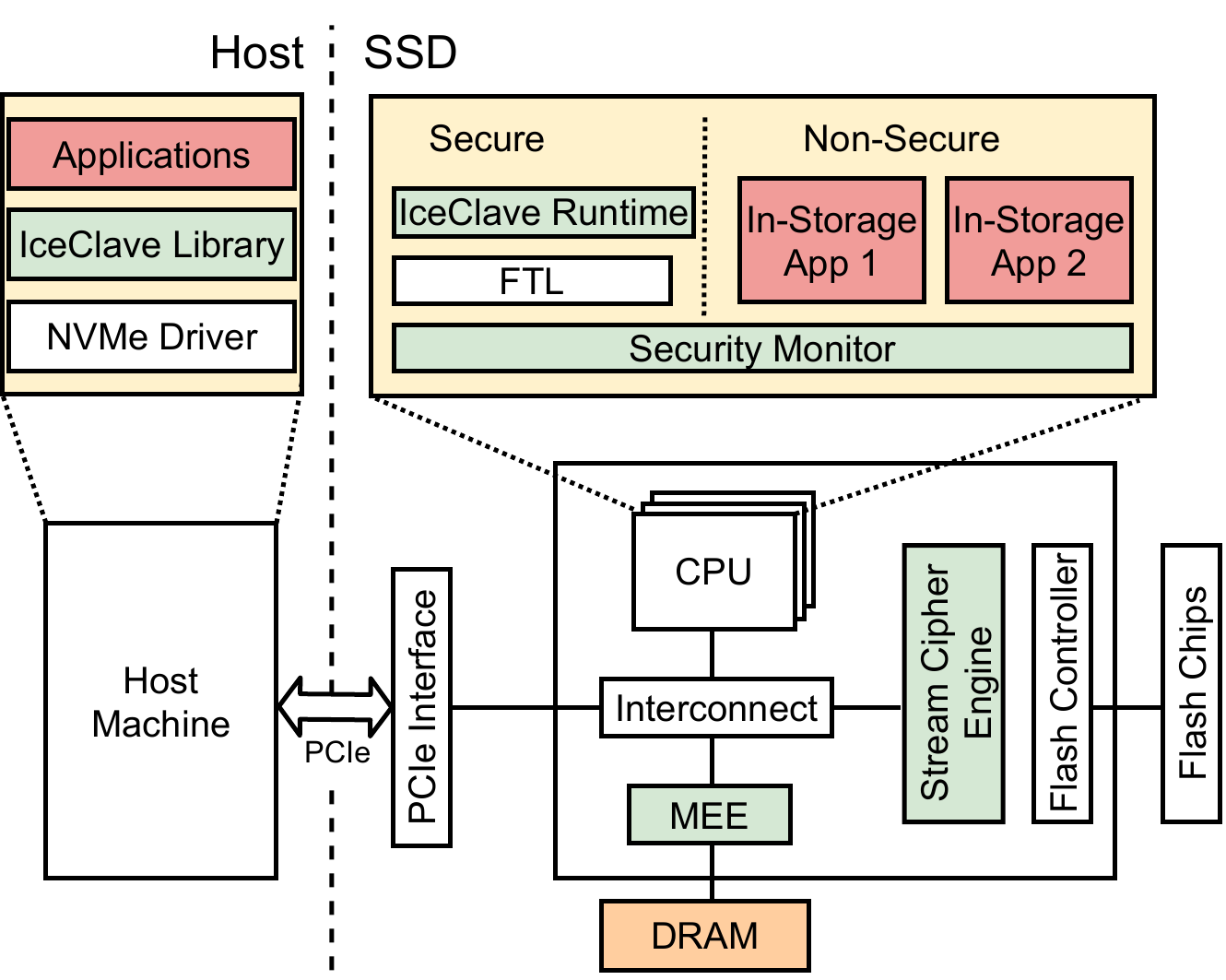}
	\vspace{-2ex}
	\caption{Overview of IceClave architecture.}
	\label{fig:overview}
		\vspace{-4ex}
\end{figure}

We present a TEE for in-storage computing with 
minimal performance and hardware cost.  
We show the overview of IceClave architecture in \figref{fig:overview}.
To achieve our goal, we propose to extend ARM TrustZone to create secure and normal world for 
security isolation and protection of different entities in FTL, while enabling memory encryption and 
verification with memory encryption engine (MEE). 

\subsection{Challenges of Building IceClave}
\label{sec:challenges}
To develop IceClave, we have to overcome three challenges. 

\begin{itemize}[leftmargin=*]
	\item First, as SSD is shared by multiple applications, we need to ensure proper security isolation. 
		Specifically, we need to not only enforce security isolation between in-storage applications and FTL functions, 
		but also the isolation between applications and IceClave runtime ($\S$\ref{subsec:ftl} and $\S$\ref{subsec:mapping}). 
\vspace{1ex}
    \item Second, to protect data of in-storage programs at runtime, IceClave needs to ensure 
	    the data security whenever the user data leaves the flash chips ($\S$\ref{subsec:memory}).
\vspace{1ex}
	\item Third, SSD controller has limited resource, such as DRAM capacity and processing capability; therefore, 
		IceClave should be lightweight and not significantly affect the performance of in-storage applications ($\S$\ref{subsec:runtime} and $\S$\ref{subsec:all}).
\vspace{1ex}
\end{itemize}

In the following sections, we will discuss how we address each of these challenges in details, respectively.

\subsection{Protecting Flash Translation Layer}
\label{subsec:ftl}
As FTL manages flash blocks and controls how user data is mapped to each flash page, 
its protection is crucial. 
If any malicious in-storage programs gain control over it, they can read, erase, or overwrite data from other users, 
which can cause severe consequences, such as data loss and leakage. 
And IceClave runtime manages how each in-storage application is initialized inside SSD, and maintains 
their metadata, such as in-storage program identity. 
If any in-storage program gains access to the metadata, the adversary can easily compromise the security of other in-storage programs. 

To protect FTL and IceClave runtime from malicious in-storage programs, we need to ensure memory 
protections for different entities in the SSD. Specifically, we have to guarantee offloaded applications 
cannot access memory regions used by FTL and IceClave runtime. We also need to ensure offloaded applications 
cannot access each other's memory regions without proper permissions.  

To achieve this, 
a straightforward way is to use TrustZone to create secure and normal worlds, and then place FTL functions and 
IceClave runtime in the secure world, and place all in-storage applications in the normal world. 
However, this will cause significant performance overhead for in-storage applications. 
This is because when an application accesses a flash page each time, it needs to context switch to the 
secure world which hosts the FTL and its address mapping table. 
Similarly, with the SGX-like approach, we can place FTL and in-storage programs in different enclaves, 
it will also generate significant performance overhead, as we have to switch frequently from one enclave to another. 

\begin{figure}
	\centering
	\includegraphics[width=0.72\linewidth]{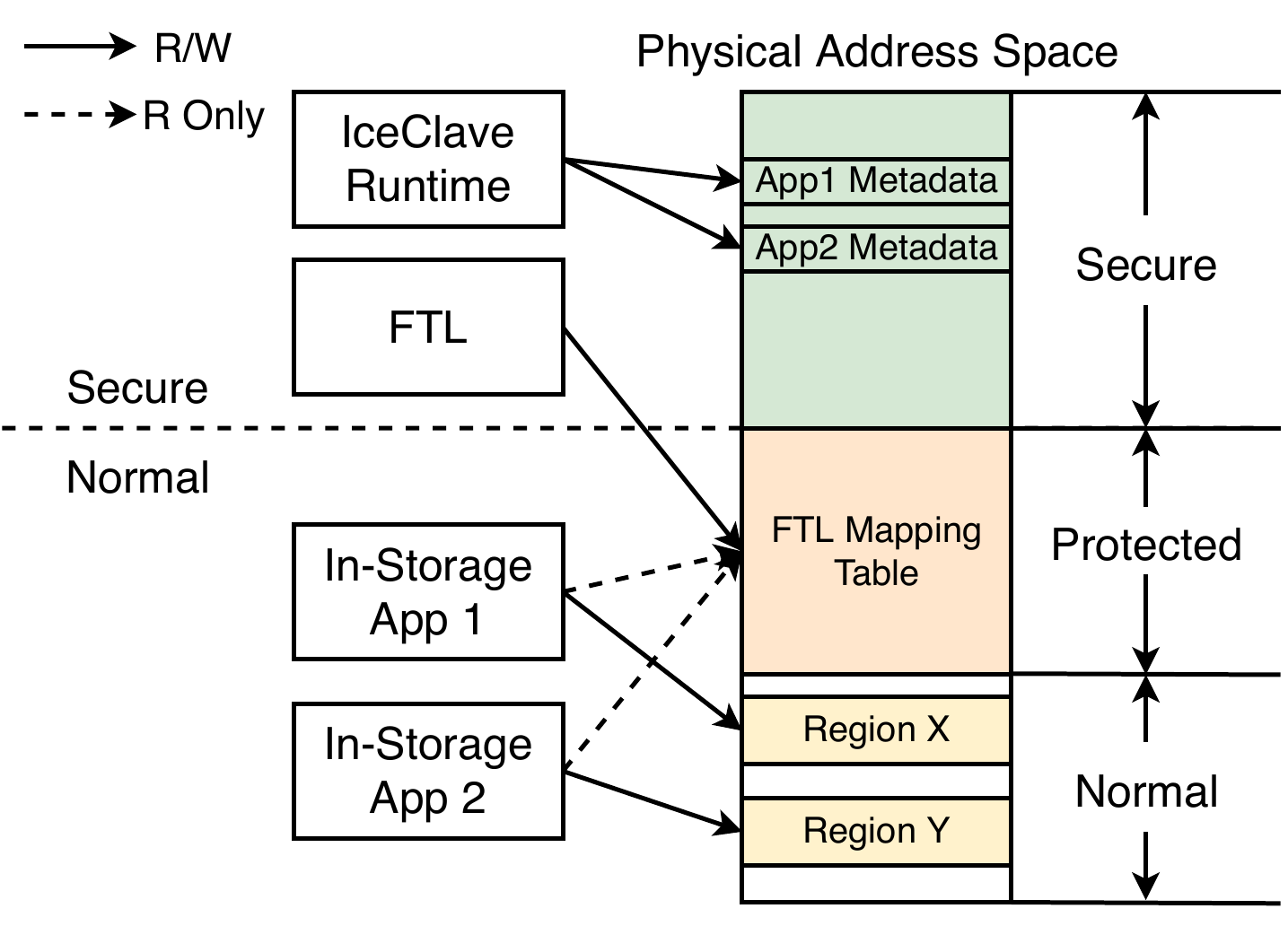}
	\vspace{-3ex}
	\caption{Memory protection regions in IceClave.
	}
	\label{fig:dram_iso}
	\vspace{-3ex}
\end{figure}

\begin{figure}
	\centering
	\includegraphics[width=0.98\linewidth]{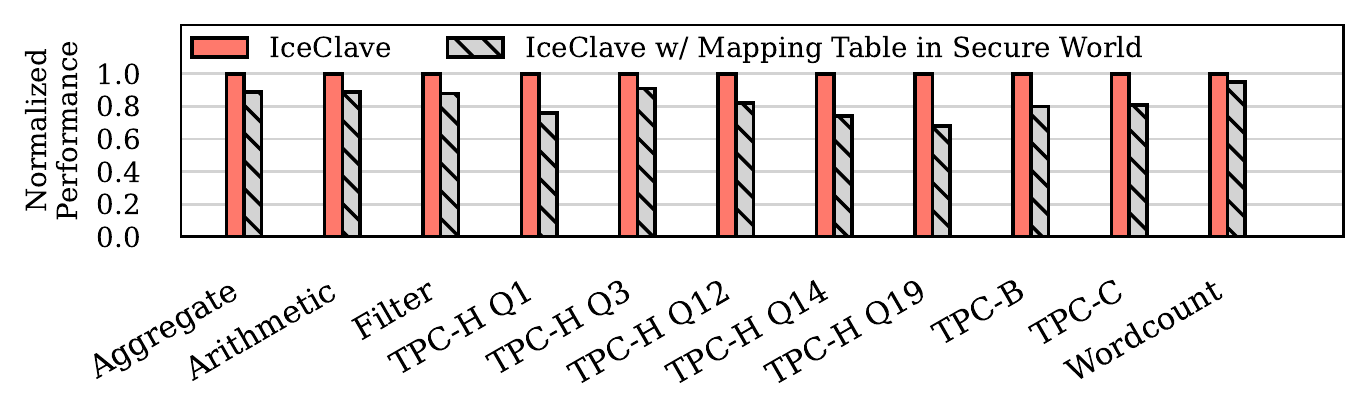}
	\vspace{-3ex} 
	\caption{Performance comparison between IceClave and IceClave with FTL mapping table in secure world. Performance is normalized to IceClave.
	}
	\label{fig:context_switch}
	\vspace{-3ex}
\end{figure}

To address this challenge, 
we partition the entire physical main memory space into 
three memory regions: normal, protected, and secure by extending TrustZone. 
We show these memory regions in Figure~\ref{fig:dram_iso}. 
Specifically, we allow FTL and IceClave runtime to execute in the secure world. And they have read/write 
permission to access the entire memory space. This is necessary, since the core functions of FTL need 
to manage the address mapping table, and 
IceClave runtime needs to manage each in-storage application, such as its TEE creation and deletion. 
We place in-storage applications in the normal world; therefore, they cannot 
access any code or data regions that belong to the FTL or IceClave runtime. 

For the protected memory region in the 
normal world, we use it to host the shared address mapping table, such that in-storage applications can only read the mapping
table entries for address translation, without paying the context-switch overhead.
Figure{~\ref{fig:context_switch}} shows this optimization can improve the performance 
of in-storage applications
by 21.6\% on average, compared to the scheme with the FTL mapping table in the secure world.
We will discuss the details of the 
flash access procedure and associated protection in $\S$~\ref{subsec:all}. 

\begin{figure}[t]
	\centering
	\includegraphics[width=0.9\linewidth]{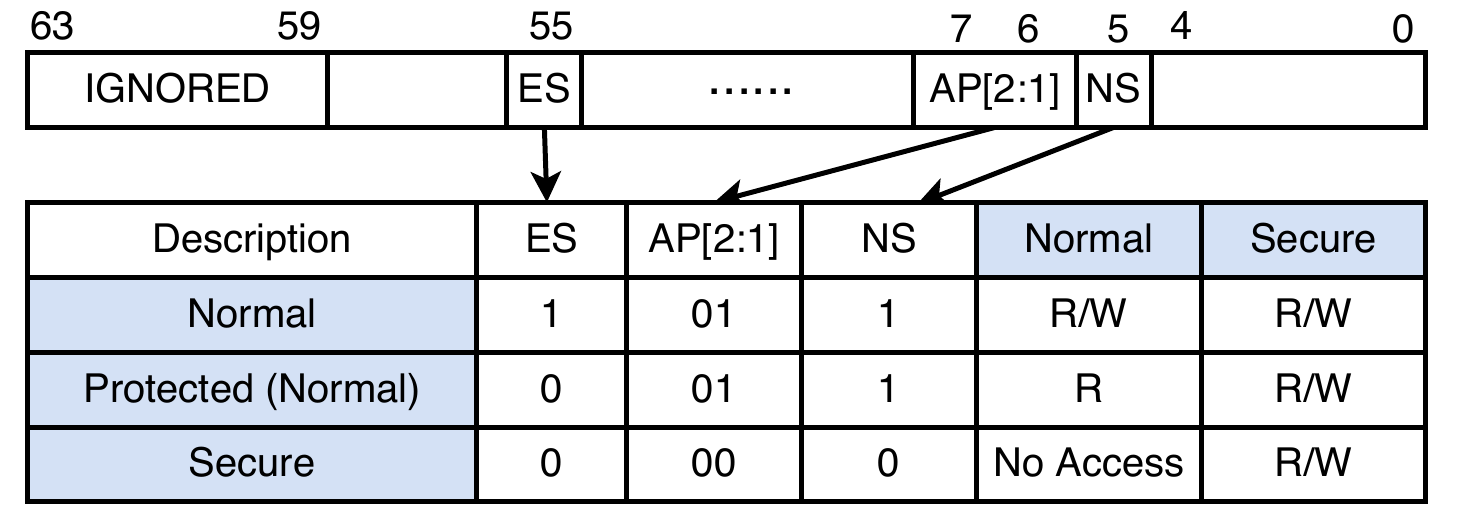}
	\vspace{-2ex}
	\caption{In-storage memory protection in IceClave.
	}
	\label{fig:mmu}
	\vspace{-3ex}
\end{figure}

We demonstrate the details of the memory region attributes in \figref{fig:mmu}.   
Following the MMU specification of ARMv8~\cite{armv8}, we use the non-secure (NS) bit to indicate whether the memory access is performed with secure or normal right.  
We utilize the access control flags (AP[2:1]) and a reserved bit (ES bit in Figure~\ref{fig:mmu}) to create the protected region, in which IceClave gives 
read-only permission to the normal world and read/write permission to the secure world. 
It is worth noting that the in-storage memory protection can be easily implemented in an older version of ARM processors~\cite{arm1156} by 
specifying the access control flags AP[2:0] as well as other processors such as RISC-V (see the discussion in $\S$\ref{subsec:discussion}).

\subsection{Access Control for In-Storage Programs}
\label{subsec:mapping}
Although each in-storage program only has the read access permission when accessing the mapping table of 
the FTL, a malicious in-storage program 
could probe the mapping table entries (e.g., by brute-force)
that are managing the address translation for the data belonging to other 
in-storage programs. Henceforth, adversaries can easily access the data of other programs. 

To address this challenge, we extend the address mapping table of FTL.  
We use the ID bits in each entry (8 bytes per entry) to track the identification of each in-storage TEE, 
and use them to verify whether an in-storage TEE has the permission to access the mapping table entry or not. 
The permission checking of flash accesses is performed with a dedicated process. It receives 
flash access requests from in-storage applications and performs permission checks before issuing the requests 
to the flash chips. This process has exclusive access to the flash chips in the normal world, which 
prevents unauthorized flash accesses from a malicious in-storage program.
We use four bits for the ID by default, which
introduces small (6.25\%) storage cost to the mapping table. IceClave will reuse the ID for newly created TEEs within 
its runtime, and set the ID bits in the mapping table upon TEE creation (see the details in $\S$\ref{subsec:all}).  

Each in-storage program only has accesses to the address mapping table of the FTL and allocated memory space. 
Accesses to other memory locations will result in a fault in the memory management unit. 
To further enhance the memory protection for in-storage programs, we also enable memory encryption and 
verification. 
\subsection{Securing In-Storage DRAM}
\label{subsec:memory}

In-storage programs load data from flash chips to the SSD DRAM for data processing. 
To conceal the data read from flash chips, IceClave
secures the data transfer procedure by encrypting the accessed data before it is transmitted on the internal bus.
Modern SSDs have employed dedicated encryption engine~\cite{www:encryptssd}, however, it is a cryptography co-processor mainly used for full-disk 
encryption. In this work, we develop a lightweight stream cipher engine in the SSD controller for securing the data transfers from flash chips to 
the storage processor (see the implementation details in $\S$\ref{sec:implt}).

Although we enable the data encryption as we transfer data between SSD DRAM and flash controllers, 
the user data that includes raw data, intermediate data, and produced results could still 
be leaked at runtime. 
To address this challenge, IceClave enables both memory encryption and integrity verification. 

\textbf{Memory Encryption.}
The goal of memory encryption is to protect any data or code in a memory access from being leaked. 
To achieve this, a common approach is to encrypt the cache lines in the processor, when they are being written to memory. 
The state-of-the-art work usually uses split-counter encryption~\cite{encryption:isca06, vault:asplos18, triad:isca19}. 
It works by encrypting a cache line through an XOR with a pseudo one time pad (OTP), 
and OTP is generated from encrypting a counter through a block cipher such as AES. 
The counter is incremented after each write back to guarantee temporal uniqueness.  
It is encoded as a concatenation of a major counter and a minor counter. When a minor counter overflows, 
the major counter is incremented, and all other minor counters are reset. The associated memory blocks also need to be re-encrypted. 
Therefore, such an encryption scheme has significant performance overhead.

\begin{table}
	\centering
	\caption{In-storage workload characterization.}
	\vspace{-2ex}
	\footnotesize
	\label{tab:memfootprint}
	\begin{tabular}{|c|c|c|c|}
		\hline
		\textbf{Workload} & \textbf{Write Ratio} & \textbf{Workload} & \textbf{Write Ratio}\\ 
		\hline
 		Arithmetic      & \num{2.02e-4}  & TPC-H Query 1   & \num{6.40e-6}     \\ 
		Aggregate       & \num{2.08e-4}  & TPC-H Query 3   & \num{3.96e-3}    \\ 
		Filter          & \num{1.71e-4}  & TPC-H Query 12  & \num{2.99e-5} \\ 
		TPC-B           & \num{5.19e-2} & TPC-H Query 14  & \num{3.94e-6}  \\ 
		TPC-C           & \num{9.05e-2} & TPC-H Query 19  & \num{9.92e-7}   \\ 
		Wordcount       & \num{4.61e-1} & & \\
		\hline
	\end{tabular}
		\vspace{-3ex}
\end{table}

This is less of a concern for in-storage computing because it is read intensive. 
We conduct a study of typical in-storage applications (see Table~\ref{tab:workload}), and profile their  
number of memory accesses when running them (see the experimental setup in $\S$\ref{subsec:setup}). 
Our observation is that most of these applications have a trivial portion of memory writes (see Table~\ref{tab:memfootprint}). 
These writes are usually caused by the produced intermediate data of in-storage programs at runtime. Based on this observation, we design a hybrid-counter scheme.

\textbf{Hybrid-Counter Scheme.} 
The key idea is that we only use major counters for read-only pages, and for writable pages, we apply the traditional split-counter scheme.
As the minor counters will not change as long as the pages are read-only, we do not need minor counters for read-only pages.
In this case, we can improve the caching performance for in-storage applications by packing more counters (eight for read-only pages) per cache line. 

\begin{figure}[t]
	\centering
	\includegraphics[width=0.98\linewidth]{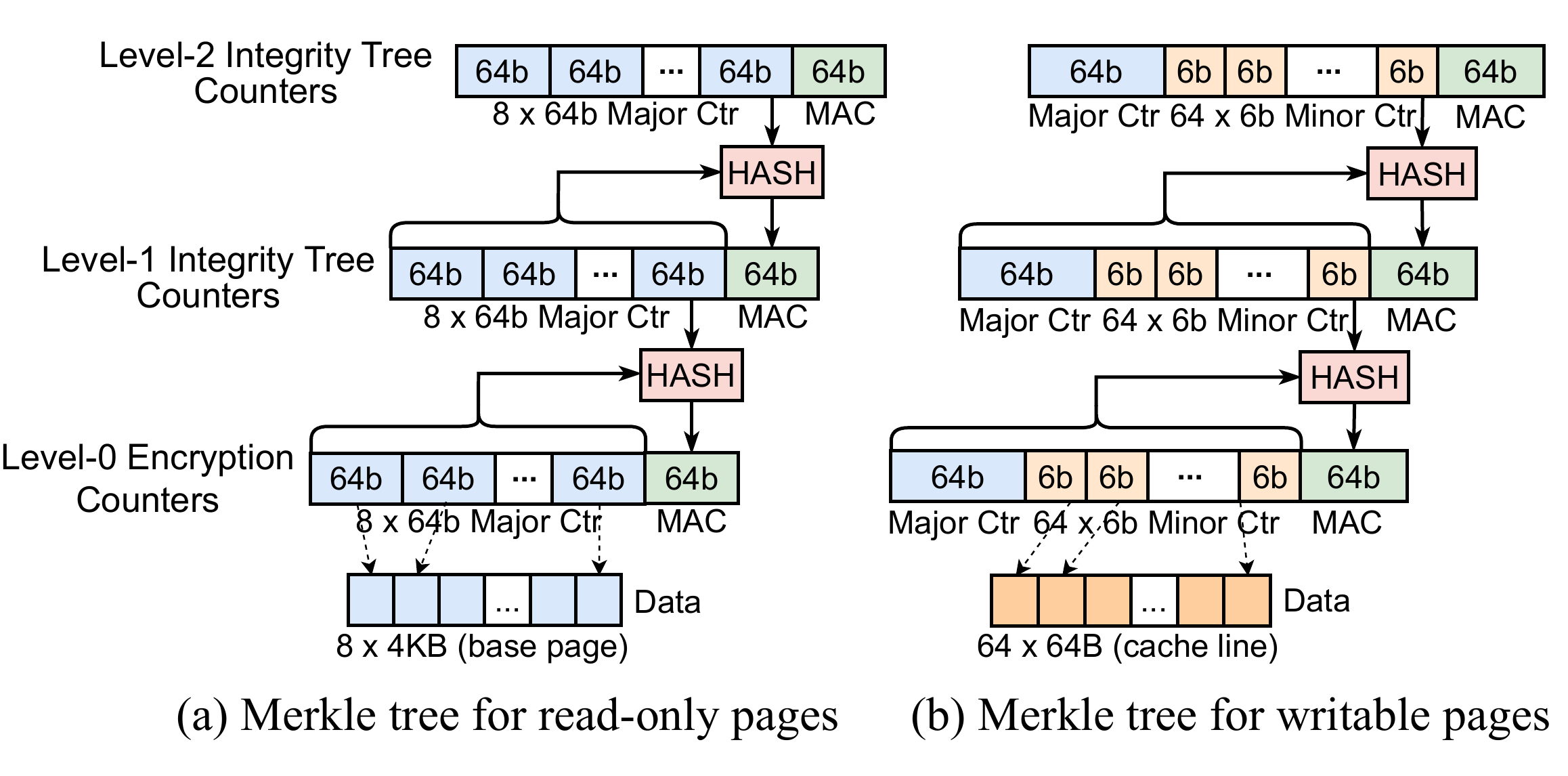}
	\vspace{-3ex}
	\caption{Memory encryption and integrity trees in IceClave.
	}
	\label{fig:bmt}
		\vspace{-2ex}
\end{figure}

\begin{figure}[t]
	\centering
	\includegraphics[width=\linewidth]{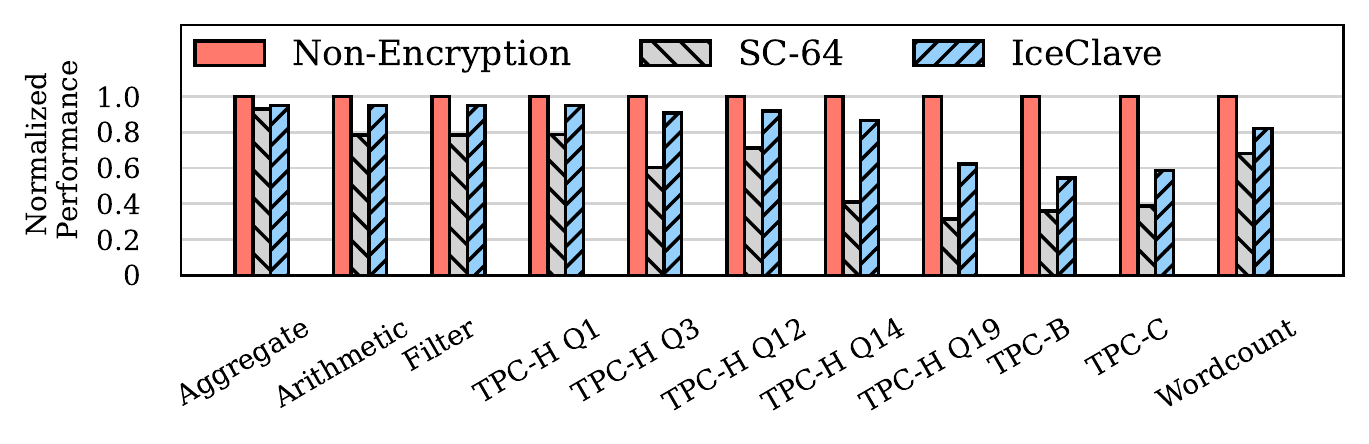}
	\vspace{-5.5ex} 
	\caption{Performance comparison of Non-encryption, Split Counters (SC-64), and IceClave. It is normalized to the scheme  
	without memory encryption.
	}
	\label{fig:ipc}
	\vspace{-2ex}
\end{figure}

The hybrid-counter scheme maintains two types of counter blocks: 
split-counter blocks for writable pages, and major-counter blocks for read-only pages, as shown in Figure~\ref{fig:bmt}.
We use two integrity trees to store the two types of counter blocks respectively. Although this requires slightly more 
memory space (0.01\% of 4GB DRAM capacity)\footnote{Given a 4GB DRAM, IceClave requires 0.5MB for the merkle tree used in
Figure~\ref{fig:bmt}(a), and 4MB for the merkle tree used in Figure~\ref{fig:bmt}(b).} to store the integrity trees, 
and needs two processor registers to store the two root message authentication codes (MACs) for integrity verification, the hybrid-counter scheme delivers 
improved performance by 43\% on average (see Figure~\ref{fig:ipc}) for in-storage programs, compared to the current split-counter scheme. 

\begin{table*}[t]
    \centering
	\caption{IceClave API}
	\vspace{-2ex}
    \footnotesize
    \label{tab:api}
	\begin{tabular}{|p{220pt}<{\raggedright}|p{222pt}<{\raggedright}|}
        \hline
		\textbf{API in IceClave Library}    & \textbf{Description} \\ \hline
		{OffloadCode (char$\ast$ bin, uint$\ast$ lpa, void$\ast$ args, uint tid)} & Invoke an offloading procedure specified by tid. \\ \hline
		GetResult (uint tid, uint64\_t$\ast$ res) & Retrieve results from the offloaded program with tid. \\ \hline \hline 
		\textbf{API in IceClave Runtime}   & \textbf{Description} \\ \hline
    	CreateTEE (void$\ast$ config, id\_t $\&$eid, tee\_t$\ast$ TEE) & Initiate a TEE and copies specified code into the TEE. \\ \hline
    	SetIDBits (const id\_t $\&$eid, uint64\_t$\ast$ lpns) & Set ID bits of the corresponding addr. mapping table entries. \\ \hline
    	TerminateTEE (tee\_t$\ast$ TEE) & Terminate the specified TEE, and reclaim resources. \\ \hline
    	ThrowOutTEE (tee\_t$\ast$ TEE, TEE\_MSG$\ast$ sm) & Abort the execution, and return an exception. \\ \hline
    	ReadMappingEntry (id\_t $\&$eid, uint64\_t$\ast$ lpa, uint64\_t$\ast$ ppa) & Request FTL to return the physical address. \\ \hline
	\end{tabular}
		\vspace{-3ex}
\end{table*}

As we use the hybrid-counter scheme in IceClave, we utilize the read/write permission bit in the page table entries to 
decide which counter blocks should be accessed. We also support dynamic permission changes of each memory page in IceClave. 
Specifically, for a read-only page, its corresponding counter is stored in the major-counter tree.
When the page becomes writable and is updated, its corresponding major counter is incremented and copied to the corresponding entry in 
the split-counter tree, and also, the minor counters in that entry are initialized.
At the same time, the page is re-encrypted using the new split-counter entry, which will be used for later accesses. 
When a writable page becomes read-only, its corresponding major counter is incremented and copied back to the major-counter tree.
In-storage programs can use the memory protection mechanisms offered by ARM processors (see Figure~\ref{fig:mmu}) to 
update the permissions of memory pages. For example, for the memory region used to store the input for in-storage programs, its pages 
are set to be read only; for the memory region allocated for storing intermediate data, its pages are set to be writable. 


%




\textbf{Memory Integrity Verification.}
To ensure the processor receives exactly the same content as it wrote in 
the memory most recently, a MAC is generated for each memory block by hashing its data and encryption counter. 
On each memory access, the MAC is re-computed using the data and encryption 
counter, and compared against the stored MAC, such that any changes on the data or counter can be detected. 
The integrity tree also prevents replay attack which can roll back the data and MAC to their older versions. 
As shown in Figure~\ref{fig:bmt}, an integrity tree organizes MACs in a hierarchy, and the parent MAC ensures the integrity of its child MACs. 
The root of the tree is securely stored in the processor chip. 
When a cache line is written back to the memory, the merkle tree will update all the nodes on the path from the data block to the root. 

In IceClave, we employ Bonsai Merkle Tree (BMT)~\cite{bmt:micro07}. 
It generates its first-level MAC by hashing counter blocks instead of data blocks.
As discussed, IceClave maintains two Merkle trees, but the extra memory cost is negligible, compared to the traditional BMT.

\input{runtime}

\input{alltogether}

In summary, IceClave can protect in-storage computing from both software and physical attacks with low overhead:  
(1) it enables memory encryption and verification for SSD DRAM with low overhead;
(2) it protects shared FTL without frequent context switches between normal 
and secure worlds; 
(3) it protects transferred data from flash chips to SSD 
DRAM with an efficient stream cipher engine in the SSD controller. 

\subsection{Discussion and Future Work}
\label{subsec:discussion}

In this paper, we exploit TrustZone technique in ARM processors to enable the memory protection between 
in-storage programs and FTL functions. This is driven by the fact that ARM processors are available in a majority of 
modern SSD controllers. As device vendors are also considering adopting the open-source RISC-V architecture in their 
controllers~\cite{riscv-disk, riscv-samsung, riscv-ssd}, the key idea of IceClave can also be implemented with new type of processors. 
To be specific, RISC-V defines three levels of privileges, including application level, 
supervisor level, and machine level~\cite{riscv}. We can map the normal, protected, and secure memory regions (see $\S$\ref{subsec:ftl}) 
to different memory regions in RISC-V respectively.  

Beyond using the ARM and RISC-V processors to conduct in-storage computing, 
recent works also deploy hardware accelerators in SSD controllers~\cite{xsd:microworkshop2013, graphssd:isca2019, deepstore:micro2019, dsagen:isca2020,BlueDBM:Jun:2015:BAB:2872887.2750412,GraFBoost,issd:csis:2016}. 
They are also lacking the 
support of in-storage TEEs. We wish to extend IceClave to these in-storage hardware 
accelerators as future work.


%% file: runtime.tex
\subsection{IceClave Runtime}
\label{subsec:runtime}
In this section, we discuss how IceClave runtime facilitates the execution of in-storage TEEs. 
It provides the essential functions for managing in-storage TEEs, such as TEE setup, TEE lifecycle 
and metadata management, and the interaction with the secure world. 
IceClave runtime also interacts with IceClave library deployed in the host machine. Note that 
IceClave library only exposes basic offloading interfaces (e.g., RPC) to end users. 
This not only reduces the trusted computing base but also simplifies the development of in-storage programs. 
We list the APIs of IceClave in Table~\ref{tab:api}. 

IceClave allows a user to interact with the SSD using two APIs: \texttt{OffloadCode} and \texttt{GetResult}. 
Once the program is offloaded to the SSD, IceClave runtime will execute \texttt{CreateTEE()} to create 
a new TEE. 
At the same time, 
it will call \texttt{SetIDBits()} to set the ID bits (access permission, see $\S$\ref{subsec:mapping}) 
of the corresponding address mapping table entries in FTL with the list of logical page addresses specified 
by the in-storage program. According to our study on the popular in-storage programs, their code size 
is 28--528KB. However, for an offloaded program whose size is larger than the available space of SSD DRAM, 
the TEE creation will fail.
During the execution of an in-storage program, \texttt{ThrowOutTEE()} will be called to handle program 
exceptions. IceClave runtime will abort the TEE for these cases that include (1) access control is violated, 
(2) TEE memory or metadata is corrupted, and (3) in-storage program throws an exception.
Once the in-storage program is finished, IceClave runtime will call the \texttt{TerminateTEE()} to terminate 
the TEE. 

With the assistance of TrustZone, IceClave supports dynamic memory allocation within each TEE. To avoid 
memory fragmentation, IceClave will preallocate a large contiguous memory region (16MB by default). Upon 
TEE deletion, IceClave runtime will release the preallocated memory region.

%% file: alltogether.tex
\subsection{Put It All Together}
\label{subsec:all}

\begin{figure}[t]
	\centering
	\includegraphics[width=0.8\linewidth]{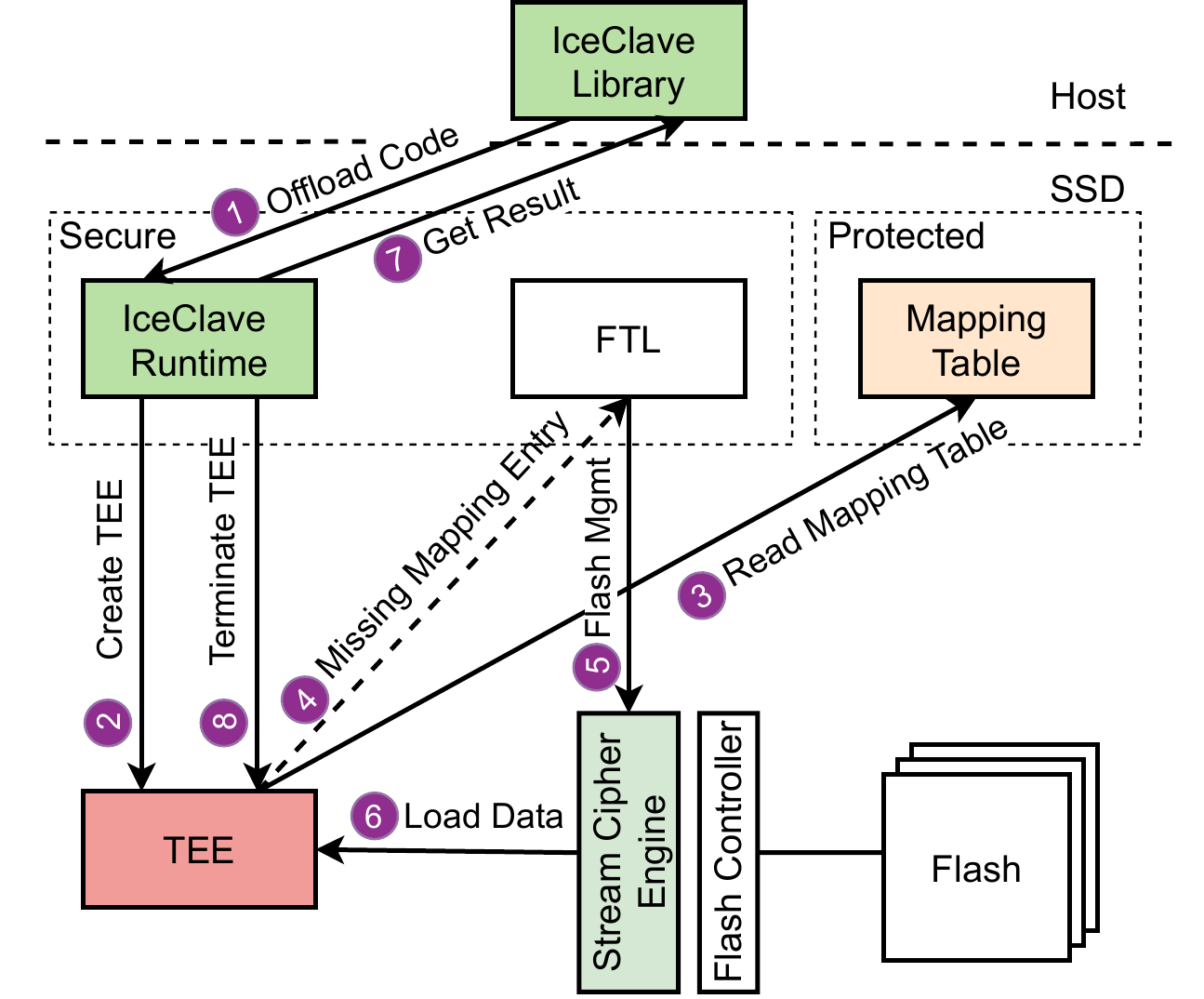}
	\vspace{-2ex}
	\caption{IceClave workflow of running in-storage programs. In this diagram, IceClave Runtime and Library, TEE, FTL, Stream Cipher Engine, and Flash Controller are compute regions. Mapping Table and Flash are memory regions.
	}
	\label{fig:workflow}
		\vspace{-3ex}
\end{figure}

We illustrate the entire workflow of running an in-storage program with IceClave in Figure \ref{fig:workflow}. 
Similar to existing in-storage computing frameworks~\cite{willow, biscuit}, IceClave library 
has a host-to-device communication layer based on PCIe, which allows users to transfer data between 
the host and SSD. As discussed in {$\S$\ref{sec:threatmodel}}, we utilize the secure channel developed in 
modern cloud computing platforms for interactions between the host and shared SSD.
The \texttt{OffloadCode} API (\circlep{1}) described in $\S$\ref{subsec:runtime} is called to offload programs. 
Its parameter \texttt{bin} represents the pre-compiled program in the form of machine code, and
\texttt{lpa} is a list of Logical Page Addresses (LPAs) of data needed by the offloaded program.
It uses task ID (\texttt{tid}) as an index for identifying the offloaded procedure.

IceClave runtime will create a new TEE for the offloaded program using \texttt{CreateTEE} ({\circlep{2}}). 
At creation, IceClave runtime will allocate memory pages from the normal memory region to the TEE, while 
the TEE metadata will be initialized and maintained in the secure memory region. IceClave also executes  
\texttt{SetIDBits} to set access permissions in the address mapping table for {LPAs}. 
The TEE does not rely on FTL to get physical page addresses (see $\S$\ref{subsec:ftl}),  
as it can access the mapping table in the protected memory region (\circlep{3}). 

However, the in-storage program may occasionally encounter cache misses, when a mapping entry for the accessed 
LPA is not cached in the SSD DRAM. In this case, the TEE has to redirect the address translation request 
to the FTL via \texttt{ReadMappingEntry} (\circlep{4}). This TEE will be paused and switched to the secure world, such 
that FTL will load the missing mapping table pages (\circlep{5}), update the cached mapping table in the protected memory 
region, and return the PPA to the TEE. To avoid in-storage programs probing the entire physical space 
of the SSD, we enforce the access control (see $\S$\ref{subsec:mapping}). 
Any data on the data path to the TEE is encrypted with the stream cipher engine (\circlep{6}). Note that 
users are encouraged to encrypt their data to defend against attacks from malicious host OS. They will send their decryption key to 
the TEE along with the offloaded program, and decrypt the data at runtime in the TEE. 
The TEE will be invoked by IceClave runtime once the in-storage program is readily prepared inside the TEE. 
IceClave runtime constantly monitors the status of initiated TEEs, secures memory regions, and ensures the 
mapping table permission guard. Exceptions will be thrown out if any aforementioned integrity is compromised. 

During the entire lifecycle of a TEE, the hardware protection mechanisms described in $\S$\ref{subsec:memory} 
are enforced to prevent physical memory attacks. IceClave will also enforce strong isolation between 
TEEs and the FTL. The memory protection described in $\S$\ref{subsec:ftl} will also be enforced in existing 
TrustZone memory controller (TZASC)~\cite{TZASC}. Once reaching the end of the TEE program, 
the results are copied into the TEE's metadata region before terminating the TEE and reclaiming used resources (\circlep{8}). 
IceClave will initiate a DMA transfer request to the host using NVMe interrupts, signaling the readiness of results. 
Results are returned to the host memory via \texttt{GetResult} (\circlep{7}) provided in IceClave library.

%% file: implt.tex
\begin{table}[t]
    \centering
    \caption{Computational SSD simulator.}
    \vspace{-2ex}
    \footnotesize
    \label{tab:simconfig}
    \begin{tabular}{|l|l|}
        \hline
    	\textbf{SSD Processor}             & ARM Cortex-A72 1.6GHz{~\cite{www:smartssd,arm:storage}}                                  \\ \hline
        Decoder Width         & 3 ops                                                                                                 \\
        Disp/Retire Width & 5 ops \\                                                                          
        L1 I/D Cache          & 48KB/32KB                                                                                             \\
        L2 Cache              & 1MB                                                                                                 \\ \hline
        \textbf{SSD DRAM}  & DDR3 1600 MHz  \\ \hline              
        	Capacity & 4GB \\
        	Organization & 1 Channel, 2 Ranks/Channel, 8 Banks/Rank \\
        	Timing & $t_{RCD}$-$t_{RAS}$-$t_{RP}$-$t_{CL}$-$t_{WR}$ = 11-28-11-11-12 \\
        	Encryption Delay      & AES-128: 60ns                                                                                         \\ \hline
        \textbf{SSD}  & 1TB Flash-based SSD \\ \hline   
        	Organization & 8 channels, 4 chips/channel, 4 dies/chip \\
        	& 2 planes/die, 2048 blocks/plane \\
        	& 512 pages/block, 4KB page\\
        	& $t_{RD}$/$t_{WR}$=50/300$\mu$s\\
        	Bandwidth & 600 MB/sec per channel \\
    	\hline	
    \end{tabular}
    	\vspace{-2ex}
\end{table}

\section{Implementation Details}
\label{sec:implt}

\noindent
\textbf{Full System Simulator.}
We implement IceClave with a computational SSD simulator developed based on the
SimpleSSD~\cite{simplessd}, Gem5~\cite{gem5}, and USIMM~\cite{usimm} simulator.
This allows us to conduct the study with different SSD configurations, which cannot be 
easily conducted with a real SSD board. 
We use the SimpleSSD to simulate a modern SSD and its storage operations. 
We show the  
SSD configuration in Table~\ref{tab:simconfig}. 
To enable in-storage computing in the simulator, we utilize Gem5 to model the 
out-of-order ARM processor in the SSD controller. 
We also implement the stream cipher in the integrated 
simulator to enable the data encryption/decryption as in-storage applications load data from flash chips. 
We use CACTI 6.5~\cite{cacti} to estimate its hardware cost, and find that the cipher engine introduces 
only 1.6\% area overhead to a modern SSD controller such as that of Intel DC P4500 SSD. 

As IceClave will enable the memory verification in SSD DRAM, we leverage USIMM to
simulate the DRAM in the SSD. We implement the Bonsai Merkle Tree (BMT) in the USIMM simulator, and 
use the hybrid-counter mode (see $\S$\ref{subsec:memory}) as the memory encryption scheme.
As discussed, the root of the integrity tree is stored in a 
secure on-chip register. The counter cache size is 128KB. 
To enable memory encryption and integrity verification, we enforce that each memory access will  
trigger the verification and update of the MAC and integrity tree. 
As we develop a full system simulator with Gem5, we run real data-intensive workloads, such as 
transaction database to evaluate the efficiency of IceClave in the following section.  

\begin{figure}[t]
        \centering
        \includegraphics[width=0.85\linewidth]{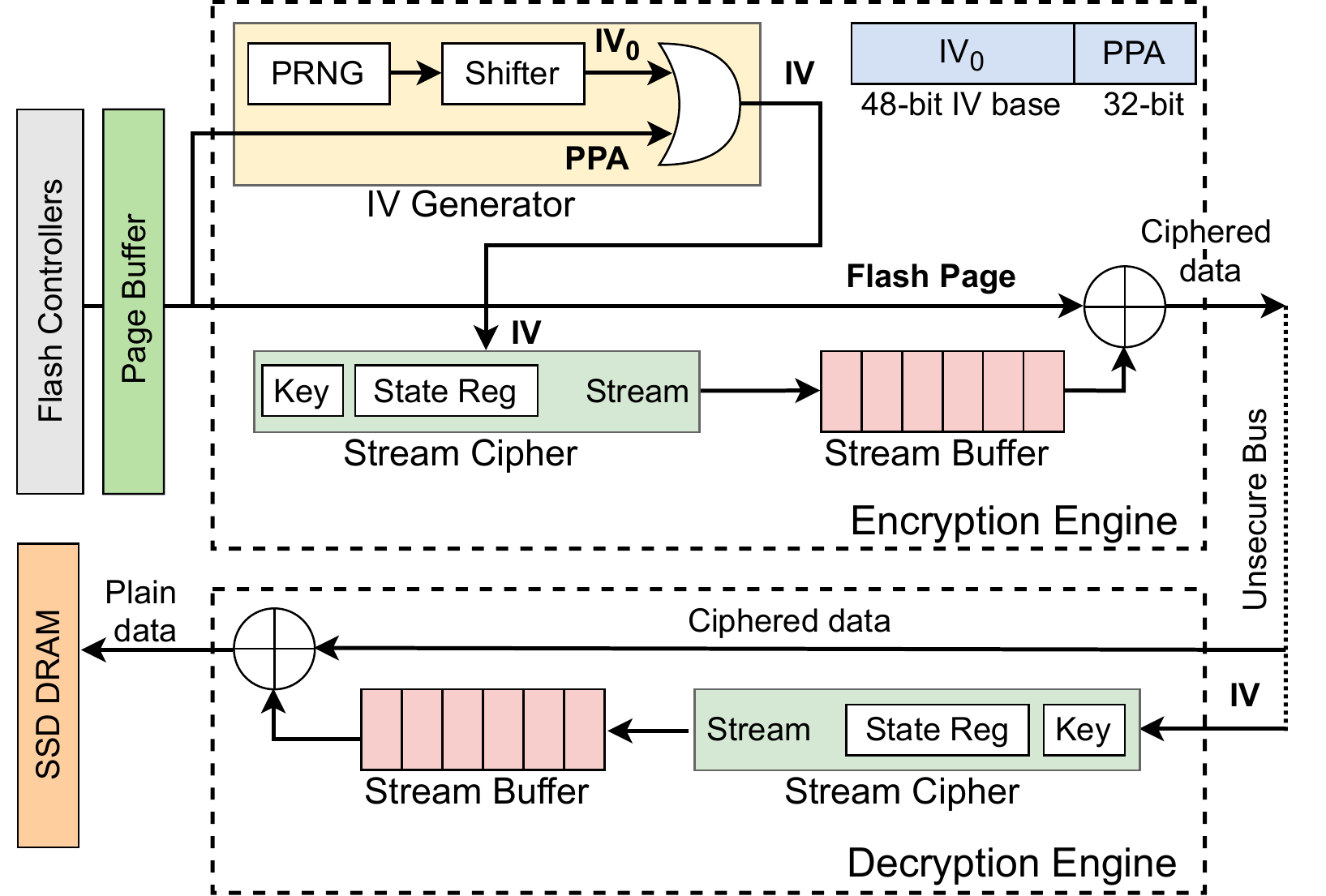}
        \vspace{-2ex}
        \caption{Stream cipher engine design in IceClave.
        }
        \label{fig:cipher_design}
        	\vspace{-2ex}
\end{figure}

\noindent
\textbf{Real System Prototype.}
{To verify the core functions of IceClave, including TEE creation/deletion, FTL, and stream cipher engine, 
we also implement IceClave with an OpenSSD Cosmos+ FPGA board} that has a Dual ARM Cortex-A9 
processor~\cite{openssd-web}. We measure their overheads and show them in Table~\ref{tab:overhead}. 
We demonstrate the architecture of the stream cipher engine in Figure~\ref{fig:cipher_design}.
Its key initialization block takes a symmetric key and an arbitrary initialization vector (IV) as the input
to initialize the cipher. IceClave keeps the key in a secure register, while the IV can be public.
Once initialized, the stream cipher generates 64 keystream bits per cycle.
The generated keystream is XORed bitwise with the data read from flash chips to produce the ciphered data.
The decipher uses the same key and IV to decode the ciphered data.
The IV is constructed with temporally unique random numbers and spatially unique address bits.
The orthogonal uniqueness enforces a strong guarantee that the same IV value will not be used twice 
during a certain period of time.
The stream cipher algorithm we used refers to the Trivium~\cite{Trivium}.
{To provide the uniqueness for different flash pages, we compose the IV by concatenating its physical page
address (PPA) and the output of a pseudo-random number generator (PRNG). }
%


%% file: evaluation.tex
\section{Evaluation}
\label{sec:eval}
Our evaluation demonstrates that: (1) IceClave introduces minimal performance overhead to in-storage workloads while
enforcing security isolation in SSD controllers ($\S$\ref{subsec:perf} and $\S$\ref{subsec:overhead}); 
(2) IceClave scales in-storage application performance as we increase the internal bandwidth
of SSDs ($\S$\ref{subsec:bandwidth}); 
(3) It can benefit various SSD devices with different access latencies ($\S$\ref{subsec:latency});
(4) IceClave still outperforms conventional host-based computing significantly while offering security isolation, 
as we vary the in-storage computing capability ($\S$\ref{subsec:computing} and $\S$\ref{subsec:dram}); 
(5) IceClave enables concurrent execution of multiple in-storage  
programs with low performance overhead ($\S$\ref{subsec:multitenant}). 

\begin{table}
    \centering
    \caption{In-storage workloads used in our evaluation.}
    \vspace{-2ex}
    \scriptsize
    \label{tab:workload}
    \begin{tabular}{|p{45pt}|p{145pt}|}
    \hline
                \textbf{Workload}   & \textbf{Description} \\ \hline
        Arithmetic & Mathematical operations against data records                                  \\ \hline
        Aggregation& Aggregate a set of values with average operation     \\ \hline
        Filter     & Filter a set of data that matches a certain feature                                        \\ \hline
        TPC-H Q1   & Query pricing summary involving scan                    \\ \hline
        TPC-H Q3   & Query shipping priority involving join              \\ \hline
        TPC-H Q12  & Query shipping modes and order priority with join                                           \\ \hline
        TPC-H Q14  & Query market response to promotion with join                                                 \\ \hline
        TPC-H Q19  & Query discounted revenue with join and aggregate           \\ \hline
        TPC-B      & Queries in a large bank with multiple branches                                  \\ \hline
        TPC-C      & Online transaction queries in a warehouse center                              \\ \hline
        Wordcount  & Count the number of words in a long text~\cite{biscuit} \\ \hline
    \end{tabular}
    	\vspace{-3ex}
\end{table}

\subsection{Experimental Setup}
\label{subsec:setup}
We evaluate IceClave with a set of synthetic workloads and real-world applications as shown in Table~\ref{tab:workload}.
In the synthetic workloads, we use several essential operators in database system, including arithmetic, aggregation, and filter operations.
As for the real-world applications, we run real queries from the TPC-H benchmark. Specifically, we use TPC-H Query 1,
3, 12, 14, and 19 that include the combination of multiple join and aggregate operations.
In addition to these workloads, we also run write-intensive workloads TPC-B, TPC-C, and Wordcount to 
further evaluate the encryption and integrity verification overhead of IceClave.
In all these workloads, we populate their dataset (tables) to
the size of 32GB, and place them across the channels in the SSD. 

We compare IceClave with several state-of-the-art solutions. Particularly, we compare IceClave with the Intel SGX available in the host machine, in which we
load the data from the SSD to the host memory and conduct the queries in the SGX. For this setting, we use a real server, which has an Intel {i7-7700K}
processor running at 4.2GHz, 16GB DDR4-3600 DRAM, and 1TB Intel DC P4500 SSD. {For fair comparison, we follow the specification of the Intel DC P4500 SSD
to configure our SSD simulator. We also compare IceClave with current in-storage computing approaches that do not provide TEEs for offloaded programs. }
We list them as follows:

\begin{itemize}[leftmargin=*]
\vspace{1ex}
    \item \textbf{Host}: in which we load data from the SSD to the host memory, and execute the data queries using host processors.
        The host machine and SSD setups are described above.
\vspace{1ex}
    \item \textbf{Host+SGX}: in which we run data queries within the Intel SGX after loading data from the SSD. The version of the
            SGX SDK we use in our experiments is 2.5.101.
\vspace{1ex}
    \item \textbf{In-Storage Computing (ISC):} in which we run data queries with the ARM processors
            in the SSD controller, such that we can exploit the high internal bandwidth of the SSD.
\vspace{1ex}
\end{itemize}


\begin{figure}[t]
    \centering
    \includegraphics[width=0.98\linewidth]{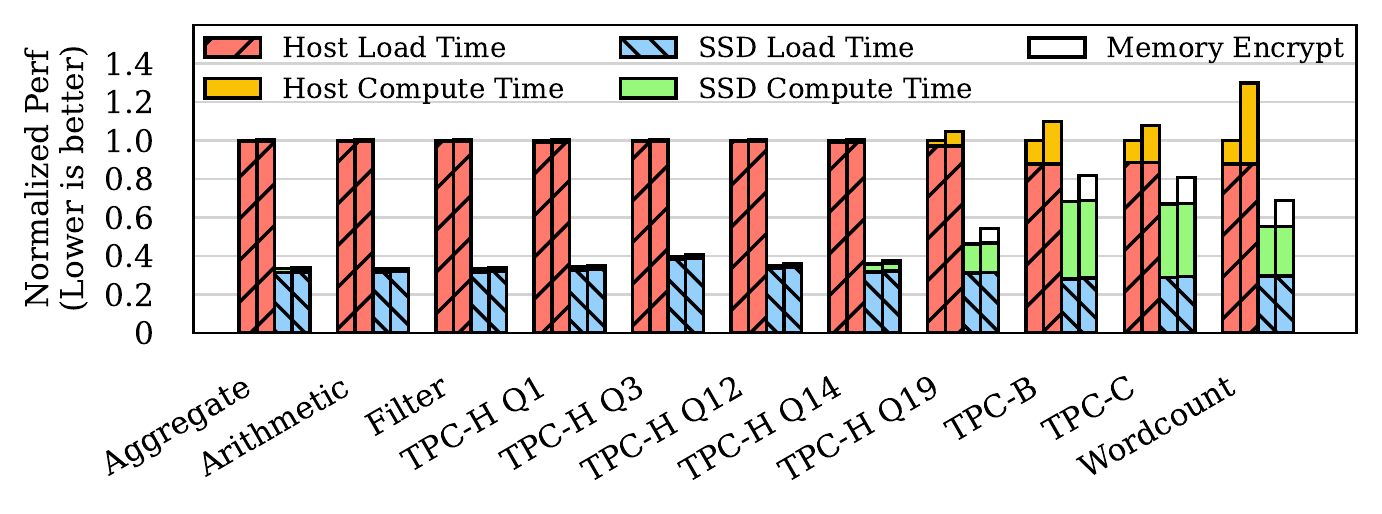}
    \vspace{-3ex}
    \caption{Performance comparison of Host, Host+SGX, ISC, and IceClave (from left to right).
    We show the performance breakdown of each scheme.}
    \label{fig:iscvsvanila}
    	\vspace{-3ex}
\end{figure}

\subsection{Performance of IceClave}
\label{subsec:perf}
We show the normalized performance of running each query benchmark in Figure~\ref{fig:iscvsvanila}.
We use the Host as the baseline, in which we run the query workload with the host machine while 
loading the dataset from the SSD. As shown in Figure~\ref{fig:iscvsvanila}, IceClave outperforms 
Host and Host+SGX by 2.31$\times$ and 2.38$\times$ on average, respectively. This shows that 
IceClave will not compromise the performance benefits of in-storage computing. 
As those data query workloads are bottlenecked by the storage 
I/O, the SGX on the host machine (Host+SGX) slightly decreases the workload performance. 
Compared to in-storage computing without security isolation enabled (ISC), IceClave introduces 
7.6\% performance overhead, due to the security techniques used in the in-storage TEE. 

To further understand the performance behaviors of IceClave, we also demonstrate the performance 
breakdown in Figure~\ref{fig:iscvsvanila}. As for the Host and Host+SGX schemes, we partition their 
workflow into two major parts: data load and computing time. As we can see, 
Host+SGX incurs 103\% extra computing time on average, caused by the SGX running in the host machine. 
As for the ISC and IceClave schemes, we profile the data load time from flash chips, the computing 
time with in-storage processors, and the overhead caused by the memory encryption and verification 
for IceClave. As shown in Figure~\ref{fig:iscvsvanila}, IceClave and ISC take much less time on loading 
time, as the internal bandwidth of the SSD is higher than its external bandwidth. And they require  
more time (2.47$\times$ on average) to execute the data queries. Compared to ISC, IceClave needs memory 
encryption and verification. 
For write-intensive workloads such as Wordcount, IceClave slightly increases memory encryption overhead. This is because Merkle tree intrinsically supports parallel updates, and our hybrid-counter design preserves this property by default.
However, IceClave still 
outperforms host-based approaches significantly for a majority of in-storage workloads. 

\begin{table}[t]
    \centering
	\caption{Overhead source of IceClave.}
	\vspace{-2ex}
    \footnotesize
    \label{tab:overhead}
	\begin{tabular}{|p{78pt}<{\raggedleft}|p{55pt}<{\raggedright}|}
        \hline
    	\textbf{Overhead Source}   & \textbf{Average Time} \\ \hline
        TEE creation &  95~$\mu$s                                 \\ \hline
        TEE deletion &   58~$\mu$s   \\ \hline
        Context switch    &  3.8~$\mu$s           \\ \hline
        Memory encryption   &    102.6~ns              \\ \hline
        Memory verification   &  151.2~ns             \\ \hline
    \end{tabular}
    	\vspace{-3ex}
\end{table}

\begin{table}[t]
	\centering
	\footnotesize
	\caption{Extra memory traffic caused by memory encryption and verification when running in-storage workloads.}
	\vspace{-2ex}
	\label{tab:memtraffic}
	\begin{tabular}{|c|c|c|}
		\hline
		\textbf{Workload} & \textbf{Encryption} & \textbf{Integrity Verification}\\ 
		\hline
 		Arithmetic & 3.05\% & 2.27\% \\
 		\hline
		Aggregate & 3.06\% & 2.26\% \\
		\hline
		Filter & 3.04\% & 2.26\% \\
		\hline
		TPC-H Query 1 & 2.99\% & 2.22\% \\
		\hline
		TPC-H Query 3 & 5.62\% & 4.5\%\\
		\hline
		TPC-H Query 12 & 5.11\% & 3.78\% \\
		\hline
		TPC-H Query 14 & 10.28\% & 5.39\%\\
		\hline
		TPC-H Query 19 & 36.20\% & 24.75\%\\
		\hline
		TPC-B & 46.92\% & 36.68\%\\
		\hline
		TPC-C & 39.09\% & 31.72\%\\
		\hline
		Wordcount & 67.45\% & 43.81\%\\
		\hline
	\end{tabular}
		\vspace{-5ex}
\end{table}

\subsection{Overhead Source in IceClave}
\label{subsec:overhead}
We also profile the entire workflow of running an in-storage program with IceClave.  
We show the critical components of IceClave and their overhead in Table~\ref{tab:overhead}. 
IceClave takes 95~$\mu$s and 58$~\mu$s (measured in real SSD FPGA board) to create and 
delete a TEE inside SSD, respectively. The overhead of context switch between 
secure world and normal world is 3.8~$\mu$s. As discussed in $\S$\ref{subsec:ftl}, 
IceClave has infrequent context switches at runtime, as it places the frequently accessed 
address mapping table in the protected memory region. The context switch happens mostly because 
the mapping table entries are missing in the protected memory region, and IceClave needs to switch 
to the secure world to fetch the mapping table from flash chips, and update 
them in the protected memory region. 

Moreover, IceClave incurs much less memory encryption and verification operations, 
because most in-storage workloads are read intensive (see $\S$\ref{subsec:memory} and Table~\ref{tab:memfootprint}). 
The average execution times of each memory encryption and verification take 102.6~ns and 151.2~ns, respectively.  
We profile the additional memory accesses (see Table~\ref{tab:memtraffic}) 
incurred by fetching and overflowing counters in the memory encryption and integrity verification. 
We show the extra memory traffic in percentage in Table~\ref{tab:memtraffic}, when comparing to the regular memory traffic without 
enforcing memory security. 
Memory encryption and verification increase the memory traffic by 20.26\% and 14.51\% on average, respectively. 

We also profile the number of flash address translations requested from a TEE, and find that only 0.17\% of these address 
translations are missed in the cached mapping table in the protected memory region. This indicates that in-storage programs do not incur the context 
switch from the normal world to the secure world frequently, showing that IceClave is lightweight for in-storage 
workloads.

%


\begin{figure}[t]
	\centering
	\includegraphics[width=\linewidth]{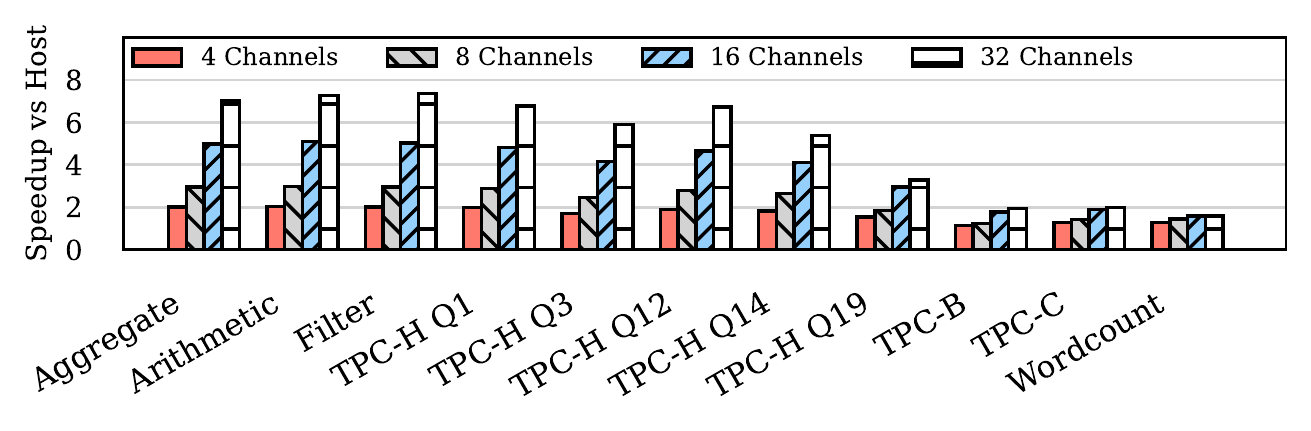}
	\vspace{-5ex} 
	\caption{IceClave scales its performance, as we vary the internal SSD bandwidth by using different number of channels (normalized to Host).}
	\label{fig:bws}
		\vspace{-4ex}
\end{figure}

\begin{figure}[t]
	\centering
	\includegraphics[width=\linewidth]{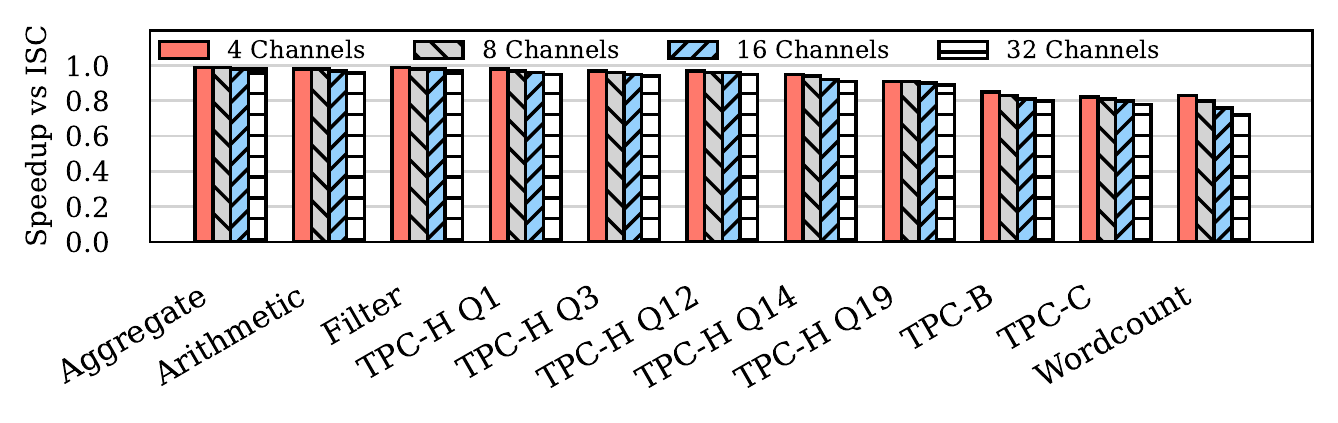}
	\vspace{-6ex} 
	\caption{IceClave introduces minimal performance overhead (normalized to ISC), as we vary the internal SSD bandwidth by using different number of channels.}
	\label{fig:bws-isc}
		\vspace{-3ex}
\end{figure}

\subsection{Impact of SSD Bandwidth}
\label{subsec:bandwidth}
We now evaluate the performance sensitivity of IceClave, as we change different SSD parameters. 
We first vary the internal bandwidth of the SSD by changing the number of flash channels from 4 to 32. With this, 
the aggregated internal I/O bandwidth grows linearly, while the external bandwidth is 
capped by the PCIe bandwidth~\cite{summarizer, lightnvm}. We compare IceClave with the host-based approach (Host), 
and present the normalized speedup in Figure~\ref{fig:bws}. 
For each data query workload, 
the performance benefit of IceClave scales significantly, as we increase the number of channels. To be specific, 
IceClave speeds up the performance by 1.7--5.0$\times$ over Host, showing that IceClave has negligible 
negative impact on the performance of in-storage computing. As for in-storage workloads that involve 
more complicated computations such as TPC-B, TPC-C, and Wordcount, increasing the internal bandwidth brings 1.2--1.8$\times$ 
performance speedup, and for others such as the synthetic workloads and TPC-H, IceClave obtain more performance benefits (1.9-6.2$\times$).    
Note that IceClave achieves even more performance benefits than Host+SGX, because SGX introduces 
extra overhead (see Figure~\ref{fig:iscvsvanila} and $\S$\ref{subsec:perf}). 
And IceClave enables TEE for in-storage programs.

{As we vary the internal SSD bandwidth, we also compare IceClave with ISC. As shown in Figure}~\ref{fig:bws-isc}, 
{IceClave decreases the application performance by up to 28\% (8.6\% on average), compared to ISC. Its 
additional overhead is slightly 
increased as we increase the number of channels for complicated data queries like TPC-C. This is mainly due to the 
increased overhead of memory encryption and integrity verification. However, IceClave offers a TEE 
for offloaded programs, making us believe it is worth the effort. }

\subsection{Impact of Data Access Latency}
\label{subsec:latency}
\begin{figure}[t]
	\centering
	\includegraphics[width=\linewidth, trim={0 0.3cm 0 0},clip]{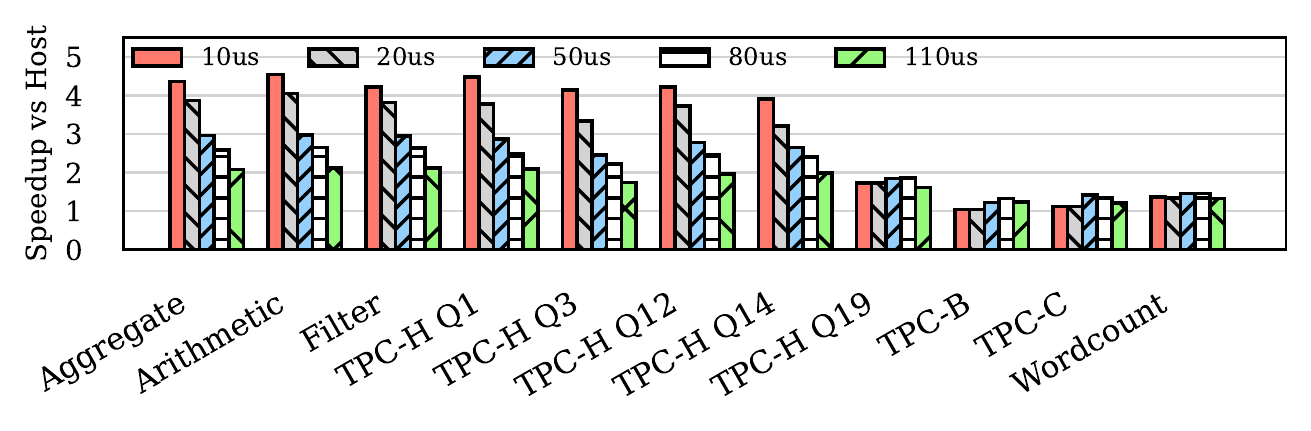}
	\vspace{-3.8ex} 
	\caption{IceClave outperforms host-based computing for I/O-intensive workloads, as we vary the flash device latency.}
	\label{fig:lat}
		\vspace{-3ex}
\end{figure}

To understand how the data access latency affects the performance of IceClave, 
{we vary the read latency of accessing a flash page from 10~$\mu$s, modeling an ultra-low 
latency NVMe SSD}~\cite{memblaze, optane_latency}, to 110~$\mu$s, modeling a commodity TLC-based SSD~\cite{microntlc}. 
We keep the write latency as 300~$\mu$s, this is because most in-storage workloads are read-intensive, 
which involve few write operations to the dataset stored 
in the SSD. We use 8 channels in the SSD. We present the experimental results in Figure~\ref{fig:lat}. 
{As shown in Figure}~\ref{fig:lat}, compared to the host-based computing approach that is bottlenecked by the 
external PCIe bandwidth, IceClave delivers performance benefit (1.8--3.2$\times$) for various 
SSD devices with different access latencies. For TPC-B, TPC-C, and TPC-H Q19 query workloads that 
require more computing resource for hash join operations, IceClave offers less performance 
benefit for the SSD with ultra-low latency, because the processors in the host machine provide more powerful computing resource. 

\subsection{Impact of Computing Capability}
\label{subsec:computing}
As we exploit embedded processors to run in-storage applications, 
it will be interesting to understand how the in-storage 
computing capability affects the efficiency of IceClave. 
We vary this parameter by using various models of embedded processor. We use our in-storage computing simulator to simulate 
the representative out-of-order (OoO) ARM processor A72, and the in-order processor A53 with different frequencies. 
We compare IceClave with the baseline Host that has an Intel i7-7700K processor running at 4.2GHz. 

\begin{figure}[t]
	\centering
	\includegraphics[width=\linewidth]{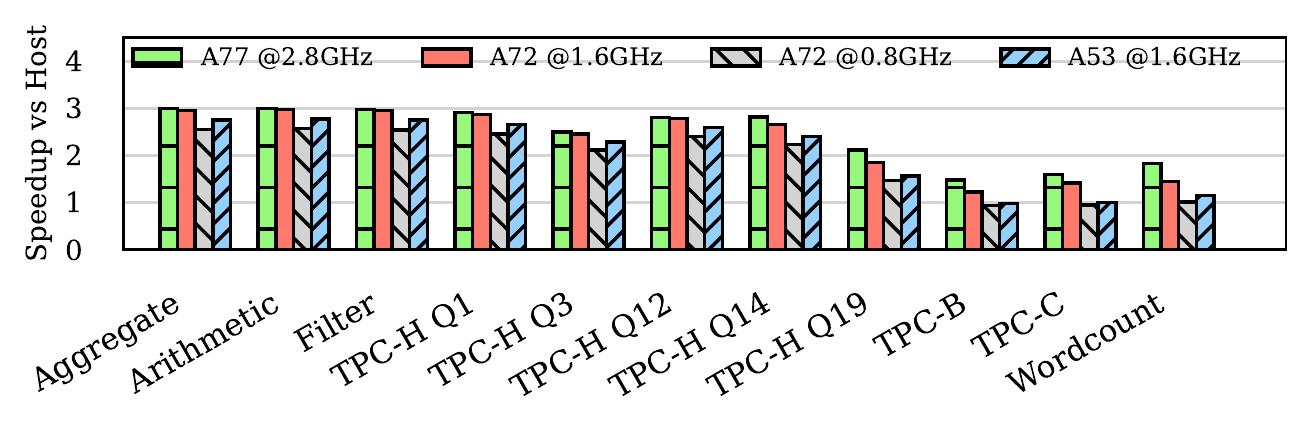}
	\vspace{-5.5ex} 
	\caption{Sensitivity analysis of IceClave, as we vary the in-storage computing capability.
	}
	\label{fig:core_ipc}
		\vspace{-3ex}
\end{figure}

We show the normalized speedup in Figure~\ref{fig:core_ipc}. The performance of IceClave drops 13.7--33.4\%
as we decrease the CPU frequency of ARM processors. And an OoO processor A72 performs slightly better than the in-order processor 
A53 with the same CPU frequency. This demonstrates that IceClave can work with different type of ARM processors and deliver 
reasonable performance benefits. 

\begin{figure}[t]
	\centering
	\includegraphics[width=\linewidth]{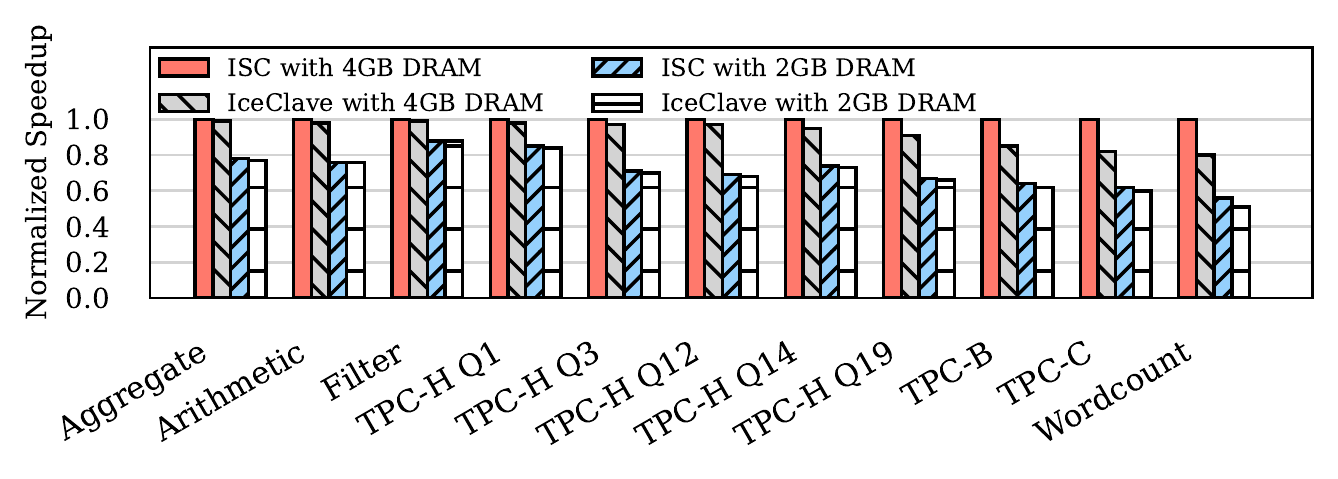}
	\vspace{-5.5ex} 
	\caption{Sensitivity analysis of IceClave, as we vary the DRAM size in the SSD controller.
	}
	\label{fig:dram-perf}
	\vspace{-3ex}
\end{figure}

\subsection{{Impact of DRAM Capacity in SSD}}
\label{subsec:dram}
{To evaluate the impact of the SSD DRAM capacity on the IceClave performance, we change the SSD DRAM size from 
4GB to 2GB while using the same configurations as described in Table}~\ref{tab:simconfig}. 
{We present the experimental results in Figure}~\ref{fig:dram-perf}. 
{As we decrease the SSD DRAM capacity, the performance of ISC drops by 12\%--44\%, as it has limited memory space to 
store its data set. The performance of IceClave follows the same trend. However, compared to ISC, IceClave still introduces 
minimal performance overhead.  
}

\begin{figure}[t]
	\centering
	\includegraphics[width=\linewidth]{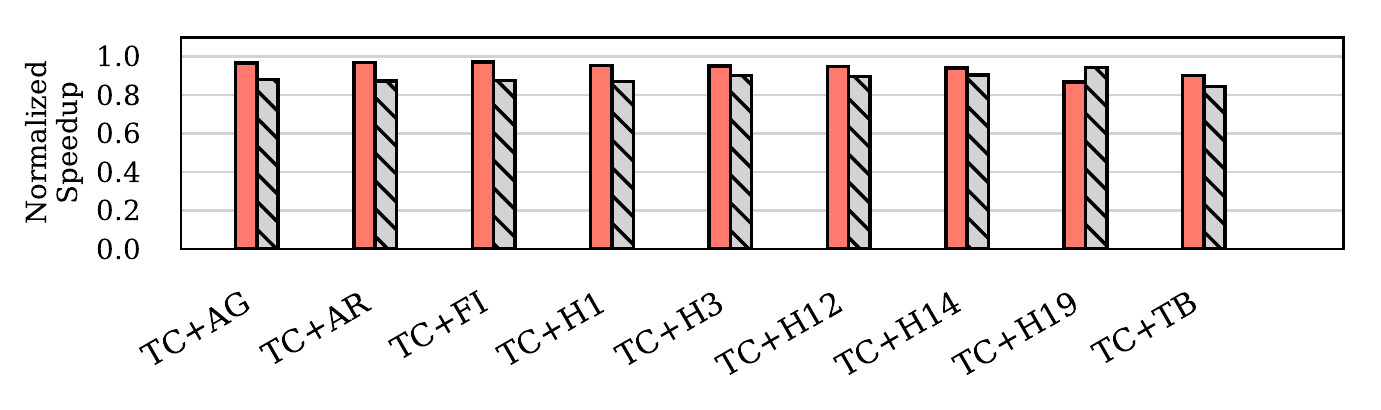}
	\vspace{-6ex} 
	\caption{IceClave performance as we run two in-storage applications concurrently.
	}
	\label{fig:twoclaves}
	\vspace{-3ex}
\end{figure}

\begin{figure}[t]
	\centering
	\includegraphics[width=\linewidth]{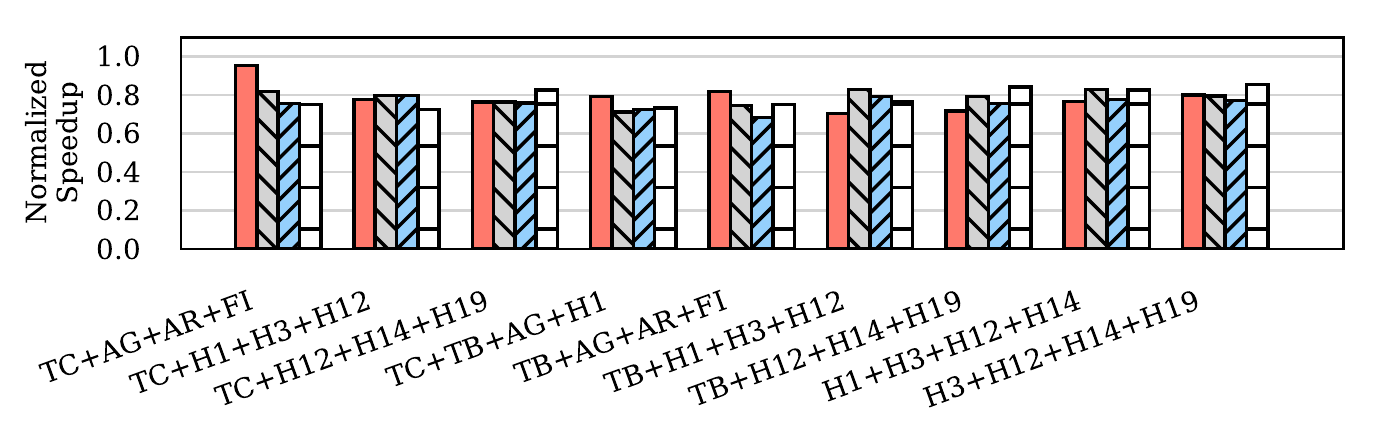}
	\vspace{-6ex} 
	\caption{IceClave performance as we run four in-storage applications concurrently.
	}
	\label{fig:fourclaves}
	\vspace{-3ex}
\end{figure}

\subsection{{Performance of Multi-tenant IceClave}}
\label{subsec:multitenant}
To further evaluate the efficiency of IceClave, 
we run multiple IceClave instances concurrently, and each instance hosts one of the 
in-storage workloads as described in Table~\ref{tab:workload}. 
We compare the application performance with the case of 
running each in-storage application independently without collocating with other instances. 
As shown in Figure~\ref{fig:twoclaves}, when we collocate the TPC-C instance with other 
workloads, the performance of in-storage applications is decreased by 6.1--15.7\%. 
As we increase the number of collocated instances (see Figure~\ref{fig:fourclaves}), 
their performance drops by 21.4\% on average. This is mainly caused by (1) the computational interference 
between the collocated IceClave instances, and (2) the increased cache misses (up to 8.7\%) 
of the cached mapping table in the protected memory region. 
However, these in-storage programs still perform better 
than host-based approaches that are constrained by the external I/O bandwidth of SSDs. 

%% file: relatedwork.tex
\section{Related Work}
\label{sec:related}

\textbf{In-storage Computing.}
In-storage computing has been intensively developed recently. 
Researchers have been exploring it for applications such as
database query~\cite{summarizer,Do:2013:QPS:2463676.2465295,kang2013enabling,smartssd},
key-value store~\cite{willow}, map-reduce workloads~\cite{biscuit,kang2013enabling},
signal processing~\cite{ActiveFlash}, and scientific
data analysis~\cite{Tiwari:2013:AFT:2591272.2591286,Tiwari:2012:RDM:2387869.2387873}.
To enable in-storage computing in modern SSDs, 
these prior works have developed various frameworks~\cite{biscuit,willow,summarizer,insider:atc19}. 
However, most of them focus on the 
programmability and performance. 
Although there is still space for improvement in these aspects, such as having SSD array and filesystem  
support for in-storage computing{~\cite{insider:atc19}}, we have to overcome the security challenge of in-storage computing for its widespread deployment, since it poses threats to user data and flash devices.
To the best of our knowledge, we are the first to propose building trusted execution environments for in-storage computing.  

\textbf{Trusted Execution Environment.}
To defend applications from malicious systems software, trusted hardware  
devices have been developed. A typical example is Intel SGX~\cite{haven}, which can create 
trusted execution environments for applications. 
{Because of the enabled security isolation, SGX is extended or customized 
to support various computing platforms}~\cite{graviton:osdi2018, scone:osdi2016, obliviate, pesos:eurosys2018, sanctum:usenixsec2016} {and 
applications}~\cite{vc3:oakland2015, enclavedb, shieldstore:eurosys2019, speicher:fast2019}. 
The hardware devices with TPM~\cite{tpm} serve the similar purpose by utilizing 
the attestation available in commodity processors from AMD and 
Intel~\cite{flicker,logicalattestation,trustvisor}. 
For ARM processors that are commonly used in 
mobile devices and storage controllers, they offer TrustZone that enables users 
to create secure world isolated from the OS~\cite{vtz, trustzone:asplos14}. Unfortunately, none of these hardware devices can be directly applied 
to in-storage computing. 
The most recent work 
ShieldStore~\cite{shieldstore:eurosys2019} {and Speicher}~\cite{speicher:fast2019} {applied SGX to key-value stores, however, none of them can protect 
the execution of in-storage programs.} 
We develop specific trusted execution environments for in-storage applications. 

\textbf{Storage Encryption and Security.} 
As we move computation closer to data in the storage devices, it would inevitably increase 
the trusted computing base, which poses security threats to user data. To protect sensitive user 
data while enabling near-data computing, a common approach is data encryption~\cite{denfs}. 
However, data leakage 
or loss would still happen at runtime, due to the lack of TEE support in 
modern SSD controllers. And adversaries can also initiate physical attacks to steal/destroy user data. 
An alternative approach is to enable computation on encrypted data~\cite{opaque:nsdi17, ocq:eurosys2020}. 
However, such an approach requires 
intensive computing resource, which cannot be satisfied by modern SSD controllers due to the 
limited resource budget~\cite{deepstore:micro2019, neardata, biscuit, timessd:eurosys2019}. 
Our work IceClave presents a lightweight approach that can enforce security isolation for in-storage 
applications as well as defend against physical attacks. 

%% file: conclusion.tex
\section{Conclusion}
\label{sec:conclusion}
Due to the lack of TEE support in SSD controllers, 
adversaries can intervene offloaded programs, mangle flash management, steal and destroy user data.
To this end, 
we develop IceClave, a
lightweight TEE which enables security
isolation between in-storage programs and flash management.
IceClave can also defend 
against physical attacks with minimal hardware cost.

%% file: ack.tex
\begin{acks}
We thank the anonymous reviewers for their comments and feedback. This work was 
partially supported by NSF grant CNS-1850317 and CCF-1919044, a gift fund from SK Hynix, and the Department of Defense under Contract FA8075-14-D-0002-0007, TAT 15-1158.
\end{acks}

%% file: main.bbl

\begin{thebibliography}{93}


\ifx \showCODEN    \undefined \def \showCODEN     #1{\unskip}     \fi
\ifx \showDOI      \undefined \def \showDOI       #1{#1}\fi
\ifx \showISBNx    \undefined \def \showISBNx     #1{\unskip}     \fi
\ifx \showISBNxiii \undefined \def \showISBNxiii  #1{\unskip}     \fi
\ifx \showISSN     \undefined \def \showISSN      #1{\unskip}     \fi
\ifx \showLCCN     \undefined \def \showLCCN      #1{\unskip}     \fi
\ifx \shownote     \undefined \def \shownote      #1{#1}          \fi
\ifx \showarticletitle \undefined \def \showarticletitle #1{#1}   \fi
\ifx \showURL      \undefined \def \showURL       {\relax}        \fi
\providecommand\bibfield[2]{#2}
\providecommand\bibinfo[2]{#2}
\providecommand\natexlab[1]{#1}
\providecommand\showeprint[2][]{arXiv:#2}

\bibitem[\protect\citeauthoryear{??}{f1}{2018}]%
        {f1}
 \bibinfo{year}{2018}\natexlab{}.
\newblock \bibinfo{title}{Amazon EC2 F1 Instances: Enable faster FPGA
  accelerator development and deployment in the cloud}.
\newblock
  \bibinfo{howpublished}{\url{https://aws.amazon.com/ec2/instance-types/f1/}}.
\newblock


\bibitem[\protect\citeauthoryear{??}{opt}{2018}]%
        {optane_latency}
 \bibinfo{year}{2018}\natexlab{}.
\newblock \showarticletitle{Intel® Optane™ SSD DC P4801X Series}.
\newblock  (\bibinfo{year}{2018}).
\newblock


\bibitem[\protect\citeauthoryear{??}{tpm}{2020}]%
        {tpm}
 \bibinfo{year}{2020}\natexlab{}.
\newblock \bibinfo{title}{TPM 2.0 Library Specification}.
\newblock
  \bibinfo{howpublished}{\\\url{https://trustedcomputinggroup.org/resource/tpm-library-specification/}}.
\newblock


\bibitem[\protect\citeauthoryear{Abulila, Mailthody, Qureshi, Huang, Kim,
  Xiong, and Hwu}{Abulila et~al\mbox{.}}{2019}]%
        {flatflash:asplos2019}
\bibfield{author}{\bibinfo{person}{Ahmed Abulila}, \bibinfo{person}{Vikram~S
  Mailthody}, \bibinfo{person}{Zaid Qureshi}, \bibinfo{person}{Jian Huang},
  \bibinfo{person}{Nam~Sung Kim}, \bibinfo{person}{Jinjun Xiong}, {and}
  \bibinfo{person}{Wen-mei Hwu}.} \bibinfo{year}{2019}\natexlab{}.
\newblock \showarticletitle{FlatFlash: Exploiting the Byte-Accessibility of
  SSDs within A Unified Memory-Storage Hierarchy}. In
  \bibinfo{booktitle}{\emph{Proceedings of the 24th International Conference on
  Architectural Support for Programming Languages and Operating Systems
  (ASPLOS'19)}}. \bibinfo{address}{Providence, RI, USA}.
\newblock


\bibitem[\protect\citeauthoryear{Ahmad, Kim, Sarfaraz, and Lee}{Ahmad
  et~al\mbox{.}}{2018}]%
        {obliviate}
\bibfield{author}{\bibinfo{person}{Adil Ahmad}, \bibinfo{person}{Kyungtae Kim},
  \bibinfo{person}{Muhammad~Ihsanulhaq Sarfaraz}, {and}
  \bibinfo{person}{Byoungyoung Lee}.} \bibinfo{year}{2018}\natexlab{}.
\newblock \showarticletitle{OBLIVIATE: A Data Oblivious Filesystem for Intel
  SGX}. In \bibinfo{booktitle}{\emph{NDSS'18}}.
\newblock


\bibitem[\protect\citeauthoryear{Anandtech}{Anandtech}{2019}]%
        {memblaze}
\bibfield{author}{\bibinfo{person}{Anandtech}.}
  \bibinfo{year}{2019}\natexlab{}.
\newblock \bibinfo{title}{Memblaze's PBlaze5 X26: Toshiba's XL-Flash-Based
  Ultra-Low Latency SSD}.
\newblock
\newblock


\bibitem[\protect\citeauthoryear{ARM}{ARM}{2007}]%
        {arm1156}
\bibfield{author}{\bibinfo{person}{ARM}.} \bibinfo{year}{2007}\natexlab{}.
\newblock \bibinfo{title}{ARM1156T2F-S Technical Reference Manual}.
\newblock
\newblock


\bibitem[\protect\citeauthoryear{Arm}{Arm}{2013}]%
        {TZASC}
\bibfield{author}{\bibinfo{person}{Arm}.} \bibinfo{year}{2013}\natexlab{}.
\newblock \bibinfo{title}{Arm CoreLink TZC-400 TrustZone Address Space
  Controller}.
\newblock
  \bibinfo{howpublished}{\\\url{http://infocenter.arm.com/help/topic/com.arm.doc.ddi0504c/DDI0504C_tzc400_r0p1_trm.pdf}}.
\newblock


\bibitem[\protect\citeauthoryear{ARM}{ARM}{2020a}]%
        {armv8}
\bibfield{author}{\bibinfo{person}{ARM}.} \bibinfo{year}{2020}\natexlab{a}.
\newblock \bibinfo{title}{ARM Architecture Reference Manual for ARMv8-A}.
\newblock
\newblock


\bibitem[\protect\citeauthoryear{ARM}{ARM}{2020b}]%
        {arm:storage}
\bibfield{author}{\bibinfo{person}{ARM}.} \bibinfo{year}{2020}\natexlab{b}.
\newblock \bibinfo{title}{ARM Storage}.
\newblock
  \bibinfo{howpublished}{\\\url{https://www.arm.com/solutions/storage}}.
\newblock


\bibitem[\protect\citeauthoryear{Arnautov, Trach, Gregor, Knauth, Martin,
  Priebe, Lind, Muthukumaran, O{\textquoteright}Keeffe, Stillwell, Goltzsche,
  Eyers, Kapitza, Pietzuch, and Fetzer}{Arnautov et~al\mbox{.}}{2016}]%
        {scone:osdi2016}
\bibfield{author}{\bibinfo{person}{Sergei Arnautov}, \bibinfo{person}{Bohdan
  Trach}, \bibinfo{person}{Franz Gregor}, \bibinfo{person}{Thomas Knauth},
  \bibinfo{person}{Andre Martin}, \bibinfo{person}{Christian Priebe},
  \bibinfo{person}{Joshua Lind}, \bibinfo{person}{Divya Muthukumaran},
  \bibinfo{person}{Dan O{\textquoteright}Keeffe}, \bibinfo{person}{Mark~L.
  Stillwell}, \bibinfo{person}{David Goltzsche}, \bibinfo{person}{Dave Eyers},
  \bibinfo{person}{R{\"u}diger Kapitza}, \bibinfo{person}{Peter Pietzuch},
  {and} \bibinfo{person}{Christof Fetzer}.} \bibinfo{year}{2016}\natexlab{}.
\newblock \showarticletitle{{SCONE}: Secure Linux Containers with Intel {SGX}}.
  In \bibinfo{booktitle}{\emph{12th {USENIX} Symposium on Operating Systems
  Design and Implementation (OSDI'16)}}. \bibinfo{address}{Savannah, GA}.
\newblock


\bibitem[\protect\citeauthoryear{Awad, Ye, Solihin, Njilla, and Zubair}{Awad
  et~al\mbox{.}}{2019}]%
        {triad:isca19}
\bibfield{author}{\bibinfo{person}{Amro Awad}, \bibinfo{person}{Mao Ye},
  \bibinfo{person}{Yan Solihin}, \bibinfo{person}{Laurent Njilla}, {and}
  \bibinfo{person}{Kazi~Abu Zubair}.} \bibinfo{year}{2019}\natexlab{}.
\newblock \showarticletitle{Triad-nvm: Persistency for integrity-protected and
  encrypted non-volatile memories}. In \bibinfo{booktitle}{\emph{Proceedings of
  the 46th International Symposium on Computer Architecture}}.
  \bibinfo{pages}{104--115}.
\newblock


\bibitem[\protect\citeauthoryear{Bae, Kim, Kim, Oh, and Park}{Bae
  et~al\mbox{.}}{2013}]%
        {issd:csis:2016}
\bibfield{author}{\bibinfo{person}{Duck-Ho Bae}, \bibinfo{person}{Jin-Hyung
  Kim}, \bibinfo{person}{Sang-Wook Kim}, \bibinfo{person}{Hyunok Oh}, {and}
  \bibinfo{person}{Chanik Park}.} \bibinfo{year}{2013}\natexlab{}.
\newblock \showarticletitle{{Intelligent SSD: A Turbo for Big Data Mining}}. In
  \bibinfo{booktitle}{\emph{Proceedings of the 22nd ACM International
  Conference of Information Knowledge Management (CIKM'13)}}.
  \bibinfo{address}{San Francisco, CA}.
\newblock


\bibitem[\protect\citeauthoryear{Bailleu, Thalheim, Bhatotia, Fetzer, Honda,
  and Vaswani}{Bailleu et~al\mbox{.}}{2019}]%
        {speicher:fast2019}
\bibfield{author}{\bibinfo{person}{Maurice Bailleu}, \bibinfo{person}{J{\"o}rg
  Thalheim}, \bibinfo{person}{Pramod Bhatotia}, \bibinfo{person}{Christof
  Fetzer}, \bibinfo{person}{Michio Honda}, {and} \bibinfo{person}{Kapil
  Vaswani}.} \bibinfo{year}{2019}\natexlab{}.
\newblock \showarticletitle{{SPEICHER}: Securing LSM-based Key-Value Stores
  using Shielded Execution}. In \bibinfo{booktitle}{\emph{17th {USENIX}
  Conference on File and Storage Technologies (FAST'19)}}.
  \bibinfo{address}{Boston, MA}.
\newblock


\bibitem[\protect\citeauthoryear{{Balasubramonian}, {Chang}, {Manning},
  {Moreno}, {Murphy}, {Nair}, and {Swanson}}{{Balasubramonian}
  et~al\mbox{.}}{2014}]%
        {neardata}
\bibfield{author}{\bibinfo{person}{R. {Balasubramonian}}, \bibinfo{person}{J.
  {Chang}}, \bibinfo{person}{T. {Manning}}, \bibinfo{person}{J.~H. {Moreno}},
  \bibinfo{person}{R. {Murphy}}, \bibinfo{person}{R. {Nair}}, {and}
  \bibinfo{person}{S. {Swanson}}.} \bibinfo{year}{2014}\natexlab{}.
\newblock \showarticletitle{Near-Data Processing: Insights from a MICRO-46
  Workshop}.
\newblock \bibinfo{journal}{\emph{IEEE Micro}} \bibinfo{volume}{34},
  \bibinfo{number}{4} (\bibinfo{year}{2014}).
\newblock


\bibitem[\protect\citeauthoryear{Baumann, Peinado, and Hunt}{Baumann
  et~al\mbox{.}}{2014a}]%
        {heaven}
\bibfield{author}{\bibinfo{person}{Andrew Baumann}, \bibinfo{person}{Marcus
  Peinado}, {and} \bibinfo{person}{Galen Hunt}.}
  \bibinfo{year}{2014}\natexlab{a}.
\newblock \showarticletitle{Shielding Applications from an Untrusted Cloud with
  Haven}. In \bibinfo{booktitle}{\emph{11th {USENIX} Symposium on Operating
  Systems Design and Implementation ({OSDI}'14)}}.
  \bibinfo{address}{Broomfield, CO}.
\newblock


\bibitem[\protect\citeauthoryear{Baumann, Peinado, and Hunt}{Baumann
  et~al\mbox{.}}{2014b}]%
        {haven}
\bibfield{author}{\bibinfo{person}{Andrew Baumann}, \bibinfo{person}{Marcus
  Peinado}, {and} \bibinfo{person}{Galen Hunt}.}
  \bibinfo{year}{2014}\natexlab{b}.
\newblock \showarticletitle{Shielding Applications from an Untrusted Cloud with
  Haven}. In \bibinfo{booktitle}{\emph{11th {USENIX} Symposium on Operating
  Systems Design and Implementation ({OSDI}'14)}}.
  \bibinfo{address}{Broomfield, CO}.
\newblock


\bibitem[\protect\citeauthoryear{Bj{\o}rling, Gonzalez, and Bonnet}{Bj{\o}rling
  et~al\mbox{.}}{2017}]%
        {lightnvm}
\bibfield{author}{\bibinfo{person}{Matias Bj{\o}rling}, \bibinfo{person}{Javier
  Gonzalez}, {and} \bibinfo{person}{Philippe Bonnet}.}
  \bibinfo{year}{2017}\natexlab{}.
\newblock \showarticletitle{{LightNVM: The Linux Open-Channel {SSD}
  Subsystem}}. In \bibinfo{booktitle}{\emph{Proceedings of the 15th {USENIX}
  Conference on File and Storage Technologies (FAST'17)}}.
  \bibinfo{address}{Santa Clara, CA}.
\newblock


\bibitem[\protect\citeauthoryear{Boboila, Kim, Vazhkudai, Desnoyers, and
  Shipman}{Boboila et~al\mbox{.}}{2012}]%
        {ActiveFlash}
\bibfield{author}{\bibinfo{person}{S. Boboila}, \bibinfo{person}{Y. Kim},
  \bibinfo{person}{S.~S. Vazhkudai}, \bibinfo{person}{P. Desnoyers}, {and}
  \bibinfo{person}{G.~M. Shipman}.} \bibinfo{year}{2012}\natexlab{}.
\newblock \showarticletitle{{Active Flash: Out-of-core Data Analytics on Flash
  Storage}}. In \bibinfo{booktitle}{\emph{Proceedings of the IEEE 28th
  Symposium on Mass Storage Systems and Technologies (MSST'12)}}.
  \bibinfo{address}{Monterey, CA}.
\newblock


\bibitem[\protect\citeauthoryear{Cao, Liu, Cheng, Zheng, Li, Wu, Ouyang, Wang,
  Wang, Kuan, Liu, Zhu, and Zhang}{Cao et~al\mbox{.}}{2020}]%
        {polardb}
\bibfield{author}{\bibinfo{person}{Wei Cao}, \bibinfo{person}{Yang Liu},
  \bibinfo{person}{Zhushi Cheng}, \bibinfo{person}{Ning Zheng},
  \bibinfo{person}{Wei Li}, \bibinfo{person}{Wenjie Wu},
  \bibinfo{person}{Linqiang Ouyang}, \bibinfo{person}{Peng Wang},
  \bibinfo{person}{Yijing Wang}, \bibinfo{person}{Ray Kuan},
  \bibinfo{person}{Zhenjun Liu}, \bibinfo{person}{Feng Zhu}, {and}
  \bibinfo{person}{Tong Zhang}.} \bibinfo{year}{2020}\natexlab{}.
\newblock \showarticletitle{{POLARDB} Meets Computational Storage: Efficiently
  Support Analytical Workloads in Cloud-Native Relational Database}. In
  \bibinfo{booktitle}{\emph{18th {USENIX} Conference on File and Storage
  Technologies ({FAST}'20)}}. \bibinfo{address}{Santa Clara, CA}.
\newblock


\bibitem[\protect\citeauthoryear{Chatterjee, Balasubramonian, Shevgoor,
  Pugsley, Udipi, Shafiee, Sudan, Awasthi, and Chishti}{Chatterjee
  et~al\mbox{.}}{2012}]%
        {usimm}
\bibfield{author}{\bibinfo{person}{Niladrish Chatterjee},
  \bibinfo{person}{Rajeev Balasubramonian}, \bibinfo{person}{Manjunath
  Shevgoor}, \bibinfo{person}{Seth Pugsley}, \bibinfo{person}{Aniruddha Udipi},
  \bibinfo{person}{Ali Shafiee}, \bibinfo{person}{Kshitij Sudan},
  \bibinfo{person}{Manu Awasthi}, {and} \bibinfo{person}{Zeshan Chishti}.}
  \bibinfo{year}{2012}\natexlab{}.
\newblock \showarticletitle{Usimm: the utah simulated memory module}.
\newblock \bibinfo{journal}{\emph{University of Utah, Tech. Rep}}
  (\bibinfo{year}{2012}).
\newblock


\bibitem[\protect\citeauthoryear{Cheerla}{Cheerla}{2019}]%
        {cstorage}
\bibfield{author}{\bibinfo{person}{Rakesh Cheerla}.}
  \bibinfo{year}{2019}\natexlab{}.
\newblock \showarticletitle{Computational SSDs}.
\newblock \bibinfo{journal}{\emph{Storage Networking Industry Association}}
  (\bibinfo{year}{2019}).
\newblock


\bibitem[\protect\citeauthoryear{Chen, Lee, and Zhang}{Chen
  et~al\mbox{.}}{2011}]%
        {chen2011essential}
\bibfield{author}{\bibinfo{person}{Feng Chen}, \bibinfo{person}{Rubao Lee},
  {and} \bibinfo{person}{Xiaodong Zhang}.} \bibinfo{year}{2011}\natexlab{}.
\newblock \showarticletitle{Essential roles of exploiting internal parallelism
  of flash memory based solid state drives in high-speed data processing}. In
  \bibinfo{booktitle}{\emph{Proceedings of the 17th IEEE International
  Symposium on High Performance Computer Architecture (HPCA'11)}}.
\newblock


\bibitem[\protect\citeauthoryear{Cho, Jeong, Oh, and Ro}{Cho
  et~al\mbox{.}}{2013}]%
        {xsd:microworkshop2013}
\bibfield{author}{\bibinfo{person}{Benjamin~Y. Cho}, \bibinfo{person}{Won~Seob
  Jeong}, \bibinfo{person}{Doohwan Oh}, {and} \bibinfo{person}{Won~Woo Ro}.}
  \bibinfo{year}{2013}\natexlab{}.
\newblock \showarticletitle{{XSD: Accelerating MapReduce by Harnessing the GPU
  inside an SSD}}. In \bibinfo{booktitle}{\emph{Proceedings of the 1st Workshop
  on Near-Data Processing in Conjunction with the 46th IEEE/ACM International
  Symposium on Microarchitecture (WoNDP)}}. \bibinfo{address}{Davis, CA}.
\newblock


\bibitem[\protect\citeauthoryear{Costan and Devadas}{Costan and Devadas}{[n.
  d.]}]%
        {www:intelsgx}
\bibfield{author}{\bibinfo{person}{Victor Costan} {and}
  \bibinfo{person}{Srinivas Devadas}.} \bibinfo{year}{[n. d.]}\natexlab{}.
\newblock \bibinfo{title}{{Intel SGX Explained}}.
\newblock \bibinfo{howpublished}{\\\url{https://eprint.iacr.org/2016/086.pdf}}.
\newblock


\bibitem[\protect\citeauthoryear{Costan, Lebedev, and Devadas}{Costan
  et~al\mbox{.}}{2016a}]%
        {sanctum:security2016}
\bibfield{author}{\bibinfo{person}{Victor Costan}, \bibinfo{person}{Ilia
  Lebedev}, {and} \bibinfo{person}{Srinivas Devadas}.}
  \bibinfo{year}{2016}\natexlab{a}.
\newblock \showarticletitle{Sanctum: Minimal Hardware Extensions for Strong
  Software Isolation}. In \bibinfo{booktitle}{\emph{25th {USENIX} Security
  Symposium ({USENIX} Security'16)}}. \bibinfo{address}{Austin, TX}.
\newblock


\bibitem[\protect\citeauthoryear{Costan, Lebedev, and Devadas}{Costan
  et~al\mbox{.}}{2016b}]%
        {sanctum:usenixsec2016}
\bibfield{author}{\bibinfo{person}{Victor Costan}, \bibinfo{person}{Ilia
  Lebedev}, {and} \bibinfo{person}{Srinivas Devadas}.}
  \bibinfo{year}{2016}\natexlab{b}.
\newblock \showarticletitle{Sanctum: Minimal Hardware Extensions for Strong
  Software Isolation}. In \bibinfo{booktitle}{\emph{25th {USENIX} Security
  Symposium ({USENIX} Security 16)}}. \bibinfo{publisher}{{USENIX}
  Association}, \bibinfo{address}{Austin, TX}, \bibinfo{pages}{857--874}.
\newblock
\showISBNx{978-1-931971-32-4}
\urldef\tempurl%
\url{https://www.usenix.org/conference/usenixsecurity16/technical-sessions/presentation/costan}
\showURL{%
\tempurl}


\bibitem[\protect\citeauthoryear{Cowan, Beattie, Johansen, and Wagle}{Cowan
  et~al\mbox{.}}{2003}]%
        {pointerguard:security2003}
\bibfield{author}{\bibinfo{person}{Crispin Cowan}, \bibinfo{person}{Steve
  Beattie}, \bibinfo{person}{John Johansen}, {and} \bibinfo{person}{Perry
  Wagle}.} \bibinfo{year}{2003}\natexlab{}.
\newblock \showarticletitle{PointGuard: Protecting Pointers From Buffer
  Overflow Vulnerabilities}. In \bibinfo{booktitle}{\emph{Proceedings of the
  12th USENIX Security Symposium (USENIX Security'03)}}.
\newblock


\bibitem[\protect\citeauthoryear{Dave, Leung, Popa, gonzalez, and Stoica}{Dave
  et~al\mbox{.}}{2020}]%
        {ocq:eurosys2020}
\bibfield{author}{\bibinfo{person}{Ankur Dave}, \bibinfo{person}{Chester
  Leung}, \bibinfo{person}{Raluca~Ada Popa}, \bibinfo{person}{Joseph~E.
  gonzalez}, {and} \bibinfo{person}{Ion Stoica}.}
  \bibinfo{year}{2020}\natexlab{}.
\newblock \showarticletitle{Oblivious Coopetive Analytics Using Hardware
  Enclaves}. In \bibinfo{booktitle}{\emph{Proceedings of European Conference on
  Computer Systems ({EuroSys}'20)}}. \bibinfo{address}{Crete, Greece}.
\newblock


\bibitem[\protect\citeauthoryear{De~Canniere and Preneel}{De~Canniere and
  Preneel}{2005}]%
        {Trivium}
\bibfield{author}{\bibinfo{person}{Christophe De~Canniere} {and}
  \bibinfo{person}{Bart Preneel}.} \bibinfo{year}{2005}\natexlab{}.
\newblock \showarticletitle{{Trivium specifications}}. In
  \bibinfo{booktitle}{\emph{eSTREAM, ECRYPT Stream Cipher Project}}.
\newblock


\bibitem[\protect\citeauthoryear{{Delkin Industrial}}{{Delkin
  Industrial}}{2019}]%
        {www:encryptssd}
\bibfield{author}{\bibinfo{person}{{Delkin Industrial}}.}
  \bibinfo{year}{2019}\natexlab{}.
\newblock \bibinfo{title}{{Encryption and Security Development in Solid State
  Storage Devices (SSD)}}.
\newblock
  \bibinfo{howpublished}{\\\url{https://www.delkin.com/blog/encryption-and-security-development-in-solid-state-storage-devices-ssd/}}.
\newblock


\bibitem[\protect\citeauthoryear{Digital}{Digital}{2019}]%
        {riscv-disk}
\bibfield{author}{\bibinfo{person}{Western Digital}.}
  \bibinfo{year}{2019}\natexlab{}.
\newblock \bibinfo{title}{RISC-V: Accelerating Next-Generation Compute
  Requirements}.
\newblock
  \bibinfo{howpublished}{\\\url{https://www.westerndigital.com/company/innovations/risc-v}}.
\newblock


\bibitem[\protect\citeauthoryear{Do, Kee, Patel, Park, Park, and DeWitt}{Do
  et~al\mbox{.}}{2013a}]%
        {Do:2013:QPS:2463676.2465295}
\bibfield{author}{\bibinfo{person}{Jaeyoung Do}, \bibinfo{person}{Yang-Suk
  Kee}, \bibinfo{person}{Jignesh~M. Patel}, \bibinfo{person}{Chanik Park},
  \bibinfo{person}{Kwanghyun Park}, {and} \bibinfo{person}{David~J. DeWitt}.}
  \bibinfo{year}{2013}\natexlab{a}.
\newblock \showarticletitle{{Query Processing on Smart SSDs: Opportunities and
  Challenges}}. In \bibinfo{booktitle}{\emph{Proceedings of the ACM SIGMOD
  International Conference on Management of Data (SIGMOD'13)}}.
  \bibinfo{address}{New York, NY}.
\newblock


\bibitem[\protect\citeauthoryear{Do, Kee, Patel, Park, Park, and DeWitt}{Do
  et~al\mbox{.}}{2013b}]%
        {smartssd}
\bibfield{author}{\bibinfo{person}{Jaeyoung Do}, \bibinfo{person}{Yang-Suk
  Kee}, \bibinfo{person}{Jignesh~M. Patel}, \bibinfo{person}{Chanik Park},
  \bibinfo{person}{Kwanghyun Park}, {and} \bibinfo{person}{David~J. DeWitt}.}
  \bibinfo{year}{2013}\natexlab{b}.
\newblock \showarticletitle{Query Processing on Smart SSDs: Opportunities and
  Challenges}. In \bibinfo{booktitle}{\emph{Proceedings of the 2013 ACM SIGMOD
  International Conference on Management of Data (SIGMOD'13)}}.
  \bibinfo{address}{New York, NY, USA}.
\newblock


\bibitem[\protect\citeauthoryear{Ferraiuolo, Baumann, Hawblitzel, and
  Parno}{Ferraiuolo et~al\mbox{.}}{2017}]%
        {komodo:sosp2017}
\bibfield{author}{\bibinfo{person}{Andrew Ferraiuolo}, \bibinfo{person}{Andrew
  Baumann}, \bibinfo{person}{Chris Hawblitzel}, {and} \bibinfo{person}{Bryan
  Parno}.} \bibinfo{year}{2017}\natexlab{}.
\newblock \showarticletitle{Komodo: Using Verification to Disentangle
  Secure-Enclave Hardware from Software}. In
  \bibinfo{booktitle}{\emph{Proceedings of the 26th Symposium on Operating
  Systems Principles (SOSP'17)}}. \bibinfo{address}{Shanghai, China}.
\newblock


\bibitem[\protect\citeauthoryear{Foundation}{Foundation}{2017}]%
        {riscv}
\bibfield{author}{\bibinfo{person}{RISC-V Foundation}.}
  \bibinfo{year}{2017}\natexlab{}.
\newblock \bibinfo{title}{The RISC-V Instruction Set Manual}.
\newblock
  \bibinfo{howpublished}{\\\url{https://content.riscv.org/wp-content/uploads/2017/05/riscv-privileged-v1.10.pdf}}.
\newblock


\bibitem[\protect\citeauthoryear{gem5~development team}{gem5~development
  team}{2020}]%
        {gem5}
\bibfield{author}{\bibinfo{person}{gem5~development team}.}
  \bibinfo{year}{2020}\natexlab{}.
\newblock \bibinfo{title}{gem5 simulator}.
\newblock
\newblock


\bibitem[\protect\citeauthoryear{Gouk, Kwon, Zhang, Koh, Choi, Kim, Kandemir,
  and Jung}{Gouk et~al\mbox{.}}{2018}]%
        {simplessd}
\bibfield{author}{\bibinfo{person}{Donghyun Gouk}, \bibinfo{person}{Miryeong
  Kwon}, \bibinfo{person}{Jie Zhang}, \bibinfo{person}{Sungjoon Koh},
  \bibinfo{person}{Wonil Choi}, \bibinfo{person}{Nam~Sung Kim},
  \bibinfo{person}{Mahmut Kandemir}, {and} \bibinfo{person}{Myoungsoo Jung}.}
  \bibinfo{year}{2018}\natexlab{}.
\newblock \showarticletitle{Amber*: Enabling Precise Full-System Simulation
  with Detailed Modeling of All SSD Resources}. In
  \bibinfo{booktitle}{\emph{2018 51st Annual IEEE/ACM International Symposium
  on Microarchitecture (MICRO)}}. IEEE, \bibinfo{pages}{469--481}.
\newblock


\bibitem[\protect\citeauthoryear{Gu, Yoon, Bae, Jo, Lee, Yoon, Kang, Kwon,
  Yoon, Cho, Jeong, and Chang}{Gu et~al\mbox{.}}{2016}]%
        {biscuit}
\bibfield{author}{\bibinfo{person}{B. Gu}, \bibinfo{person}{A.~S. Yoon},
  \bibinfo{person}{D.~H. Bae}, \bibinfo{person}{I. Jo}, \bibinfo{person}{J.
  Lee}, \bibinfo{person}{J. Yoon}, \bibinfo{person}{J.~U. Kang},
  \bibinfo{person}{M. Kwon}, \bibinfo{person}{C. Yoon}, \bibinfo{person}{S.
  Cho}, \bibinfo{person}{J. Jeong}, {and} \bibinfo{person}{D. Chang}.}
  \bibinfo{year}{2016}\natexlab{}.
\newblock \showarticletitle{Biscuit: A Framework for Near-Data Processing of
  Big Data Workloads}. In \bibinfo{booktitle}{\emph{2016 ACM/IEEE 43rd Annual
  International Symposium on Computer Architecture (ISCA'16)}}.
  \bibinfo{address}{Seoul, Korea}.
\newblock


\bibitem[\protect\citeauthoryear{Gupta, Kim, and Urgaonkar}{Gupta
  et~al\mbox{.}}{2009}]%
        {dftl}
\bibfield{author}{\bibinfo{person}{Aayush Gupta}, \bibinfo{person}{Youngjae
  Kim}, {and} \bibinfo{person}{Bhuvan Urgaonkar}.}
  \bibinfo{year}{2009}\natexlab{}.
\newblock \showarticletitle{{DFTL: A Flash Translation Layer Employing
  Demand-based Selective Caching of Page-level Address Mappings}}. In
  \bibinfo{booktitle}{\emph{Proceedings of the 14th International Conference on
  Architectural Support for Programming Languages and Operating Systems
  (ASPLOS'09)}}. \bibinfo{address}{Washington, DC, USA}.
\newblock


\bibitem[\protect\citeauthoryear{Halfacree}{Halfacree}{2018}]%
        {riscv-ssd}
\bibfield{author}{\bibinfo{person}{Gareth Halfacree}.}
  \bibinfo{year}{2018}\natexlab{}.
\newblock \bibinfo{title}{SiFive's RISC-V cores launch in two SSD families}.
\newblock
  \bibinfo{howpublished}{\\\url{https://www.bit-tech.net/news/tech/storage/sifives-risc-v-cores-launch-in-two-ssd-families/1/}}.
\newblock


\bibitem[\protect\citeauthoryear{Hua, Gu, Xia, Chen, Zang, and Guan}{Hua
  et~al\mbox{.}}{2017}]%
        {vtz}
\bibfield{author}{\bibinfo{person}{Zhichao Hua}, \bibinfo{person}{Jinyu Gu},
  \bibinfo{person}{Yubin Xia}, \bibinfo{person}{Haibo Chen},
  \bibinfo{person}{Binyu Zang}, {and} \bibinfo{person}{Haibing Guan}.}
  \bibinfo{year}{2017}\natexlab{}.
\newblock \showarticletitle{vTZ: Virtualizing {ARM} TrustZone}. In
  \bibinfo{booktitle}{\emph{26th {USENIX} Security Symposium ({USENIX}
  Security'17)}}. \bibinfo{address}{Vancouver, BC}.
\newblock


\bibitem[\protect\citeauthoryear{Huang, Badam, Caulfield, Nath, Sengupta,
  Sharma, and Qureshi}{Huang et~al\mbox{.}}{2017}]%
        {flashblox}
\bibfield{author}{\bibinfo{person}{Jian Huang}, \bibinfo{person}{Anirudh
  Badam}, \bibinfo{person}{Laura Caulfield}, \bibinfo{person}{Suman Nath},
  \bibinfo{person}{Sudipta Sengupta}, \bibinfo{person}{Bikash Sharma}, {and}
  \bibinfo{person}{Moinuddin~K. Qureshi}.} \bibinfo{year}{2017}\natexlab{}.
\newblock \showarticletitle{{FlashBlox: Achieving Both Performance Isolation
  and Uniform Lifetime for Virtualized SSDs}}. In
  \bibinfo{booktitle}{\emph{Proceedings of the 15th Usenix Conference on File
  and Storage Technologies (FAST'17)}}. \bibinfo{address}{Santa clara, CA}.
\newblock


\bibitem[\protect\citeauthoryear{Huang, Badam, Qureshi, and Schwan}{Huang
  et~al\mbox{.}}{2015}]%
        {flashmap:isca2015}
\bibfield{author}{\bibinfo{person}{Jian Huang}, \bibinfo{person}{Anirudh
  Badam}, \bibinfo{person}{Moinuddin~K. Qureshi}, {and}
  \bibinfo{person}{Karsten Schwan}.} \bibinfo{year}{2015}\natexlab{}.
\newblock \showarticletitle{{Unified Address Translation for Memory-mapped SSDs
  with FlashMap}}. In \bibinfo{booktitle}{\emph{Proceedings of the 42nd Annual
  International Symposium on Computer Architecture (ISCA'15)}}.
  \bibinfo{address}{Portland, OR}.
\newblock


\bibitem[\protect\citeauthoryear{Jun, Wright, Zhang, Xu, and Arvind}{Jun
  et~al\mbox{.}}{2018}]%
        {GraFBoost}
\bibfield{author}{\bibinfo{person}{S. Jun}, \bibinfo{person}{A. Wright},
  \bibinfo{person}{S. Zhang}, \bibinfo{person}{S. Xu}, {and}
  \bibinfo{person}{Arvind}.} \bibinfo{year}{2018}\natexlab{}.
\newblock \showarticletitle{{GraFBoost: Using Accelerated Flash Storage for
  External Graph Analytics}}. In \bibinfo{booktitle}{\emph{Proceedings of the
  45th Annual International Symposium on Computer Architecture (ISCA'18)}}.
  \bibinfo{address}{Los Angeles, CA}.
\newblock


\bibitem[\protect\citeauthoryear{Jun, Liu, Lee, Hicks, Ankcorn, King, Xu, and
  Arvind}{Jun et~al\mbox{.}}{2015}]%
        {BlueDBM:Jun:2015:BAB:2872887.2750412}
\bibfield{author}{\bibinfo{person}{Sang-Woo Jun}, \bibinfo{person}{Ming Liu},
  \bibinfo{person}{Sungjin Lee}, \bibinfo{person}{Jamey Hicks},
  \bibinfo{person}{John Ankcorn}, \bibinfo{person}{Myron King},
  \bibinfo{person}{Shuotao Xu}, {and} \bibinfo{person}{Arvind}.}
  \bibinfo{year}{2015}\natexlab{}.
\newblock \showarticletitle{{BlueDBM: An Appliance for Big Data Analytics}}.
\newblock \bibinfo{journal}{\emph{SIGARCH Comput. Archit. News}}
  \bibinfo{volume}{43}, \bibinfo{number}{3} (\bibinfo{date}{June}
  \bibinfo{year}{2015}).
\newblock


\bibitem[\protect\citeauthoryear{Kang, Kee, Miller, and Park}{Kang
  et~al\mbox{.}}{2013}]%
        {kang2013enabling}
\bibfield{author}{\bibinfo{person}{Y. Kang}, \bibinfo{person}{Y. Kee},
  \bibinfo{person}{E.~L. Miller}, {and} \bibinfo{person}{C. Park}.}
  \bibinfo{year}{2013}\natexlab{}.
\newblock \showarticletitle{{Enabling cost-effective data processing with smart
  SSD}}. In \bibinfo{booktitle}{\emph{Proceedings of the 28th IEEE Conference
  on Mass Storage Systems and Technologies (MSST'13)}}. \bibinfo{address}{Lake
  Arrowhead, CA}.
\newblock


\bibitem[\protect\citeauthoryear{Khawaja, Landgraf, Prakash, Wei, Schkufza, and
  Rossbach}{Khawaja et~al\mbox{.}}{2018}]%
        {amorphos}
\bibfield{author}{\bibinfo{person}{Ahmed Khawaja}, \bibinfo{person}{Joshua
  Landgraf}, \bibinfo{person}{Rohith Prakash}, \bibinfo{person}{Michael Wei},
  \bibinfo{person}{Eric Schkufza}, {and} \bibinfo{person}{Christopher~J.
  Rossbach}.} \bibinfo{year}{2018}\natexlab{}.
\newblock \showarticletitle{Sharing, Protection, and Compatibility for
  Reconfigurable Fabric with AmorphOS}. In \bibinfo{booktitle}{\emph{13th
  {USENIX} Symposium on Operating Systems Design and Implementation ({OSDI}
  18)}}. \bibinfo{address}{Carlsbad, CA}.
\newblock


\bibitem[\protect\citeauthoryear{Kim, Lee, and Noh}{Kim et~al\mbox{.}}{2015}]%
        {slo:fast2015}
\bibfield{author}{\bibinfo{person}{Jaeho Kim}, \bibinfo{person}{Donghee Lee},
  {and} \bibinfo{person}{Sam~H. Noh}.} \bibinfo{year}{2015}\natexlab{}.
\newblock \showarticletitle{Towards {SLO} Complying SSDs Through {OPS}
  Isolation}. In \bibinfo{booktitle}{\emph{Proceedings of the 13th {USENIX}
  Conference on File and Storage Technologies ({FAST} 15)}}.
  \bibinfo{address}{Santa Clara, CA}.
\newblock


\bibitem[\protect\citeauthoryear{Kim, Park, Woo, Jeon, and Huh}{Kim
  et~al\mbox{.}}{2019}]%
        {shieldstore:eurosys2019}
\bibfield{author}{\bibinfo{person}{Taehoon Kim}, \bibinfo{person}{Joongun
  Park}, \bibinfo{person}{Jaewook Woo}, \bibinfo{person}{Seungheun Jeon}, {and}
  \bibinfo{person}{Jaehyuk Huh}.} \bibinfo{year}{2019}\natexlab{}.
\newblock \showarticletitle{ShieldStore: Shielded In-Memory Key-Value Storage
  with SGX}. In \bibinfo{booktitle}{\emph{Proceedings of the Fourteenth EuroSys
  Conference (EuroSys'19)}}.
\newblock


\bibitem[\protect\citeauthoryear{{Kocher}, {Horn}, {Fogh}, {Genkin}, {Gruss},
  {Haas}, {Hamburg}, {Lipp}, {Mangard}, {Prescher}, {Schwarz}, and
  {Yarom}}{{Kocher} et~al\mbox{.}}{2019}]%
        {kocher:sp2019}
\bibfield{author}{\bibinfo{person}{P. {Kocher}}, \bibinfo{person}{J. {Horn}},
  \bibinfo{person}{A. {Fogh}}, \bibinfo{person}{D. {Genkin}},
  \bibinfo{person}{D. {Gruss}}, \bibinfo{person}{W. {Haas}},
  \bibinfo{person}{M. {Hamburg}}, \bibinfo{person}{M. {Lipp}},
  \bibinfo{person}{S. {Mangard}}, \bibinfo{person}{T. {Prescher}},
  \bibinfo{person}{M. {Schwarz}}, {and} \bibinfo{person}{Y. {Yarom}}.}
  \bibinfo{year}{2019}\natexlab{}.
\newblock \showarticletitle{Spectre Attacks: Exploiting Speculative Execution}.
  In \bibinfo{booktitle}{\emph{Proceedings of the 2019 IEEE Symposium on
  Security and Privacy (Oakland'19)}}.
\newblock


\bibitem[\protect\citeauthoryear{Koo, Matam, I, Narra, Li, Tseng, Swanson, and
  Annavaram}{Koo et~al\mbox{.}}{2017}]%
        {summarizer}
\bibfield{author}{\bibinfo{person}{Gunjae Koo}, \bibinfo{person}{Kiran~Kumar
  Matam}, \bibinfo{person}{Te I}, \bibinfo{person}{H.~V. Krishna~Giri Narra},
  \bibinfo{person}{Jing Li}, \bibinfo{person}{Hung-Wei Tseng},
  \bibinfo{person}{Steven Swanson}, {and} \bibinfo{person}{Murali Annavaram}.}
  \bibinfo{year}{2017}\natexlab{}.
\newblock \showarticletitle{Summarizer: Trading Communication with Computing
  Near Storage}. In \bibinfo{booktitle}{\emph{Proceedings of the 50th Annual
  IEEE/ACM International Symposium on Microarchitecture (MICRO'17)}}.
  \bibinfo{address}{Cambridge, Massachusetts}.
\newblock


\bibitem[\protect\citeauthoryear{Krahn, Trach, Vahldiek-Oberwagner, Knauth,
  Bhatotia, and Fetzer}{Krahn et~al\mbox{.}}{2018}]%
        {pesos:eurosys2018}
\bibfield{author}{\bibinfo{person}{Robert Krahn}, \bibinfo{person}{Bohdan
  Trach}, \bibinfo{person}{Anjo Vahldiek-Oberwagner}, \bibinfo{person}{Thomas
  Knauth}, \bibinfo{person}{Pramod Bhatotia}, {and} \bibinfo{person}{Christof
  Fetzer}.} \bibinfo{year}{2018}\natexlab{}.
\newblock \showarticletitle{Pesos: Policy Enhanced Secure Object Store}. In
  \bibinfo{booktitle}{\emph{Proceedings of the Thirteenth EuroSys Conference
  (EuroSys'18)}}.
\newblock


\bibitem[\protect\citeauthoryear{Lee, Kohlbrenner, Shinde, Asanovi\'{c}, and
  Song}{Lee et~al\mbox{.}}{2020}]%
        {keystone:eurosys2020}
\bibfield{author}{\bibinfo{person}{Dayeol Lee}, \bibinfo{person}{David
  Kohlbrenner}, \bibinfo{person}{Shweta Shinde}, \bibinfo{person}{Krste
  Asanovi\'{c}}, {and} \bibinfo{person}{Dawn Song}.}
  \bibinfo{year}{2020}\natexlab{}.
\newblock \showarticletitle{Keystone: An Open Framework for Architecting
  Trusted Execution Environments}. In \bibinfo{booktitle}{\emph{Proceedings of
  the Fifteenth European Conference on Computer Systems (EuroSys'20)}}.
  \bibinfo{address}{Heraklion, Greece}.
\newblock


\bibitem[\protect\citeauthoryear{Lo, Cheng, Govindaraju, Ranganathan, and
  Kozyrakis}{Lo et~al\mbox{.}}{2015}]%
        {heracles:isca2015}
\bibfield{author}{\bibinfo{person}{David Lo}, \bibinfo{person}{Liqun Cheng},
  \bibinfo{person}{Rama Govindaraju}, \bibinfo{person}{Parthasarathy
  Ranganathan}, {and} \bibinfo{person}{Christos Kozyrakis}.}
  \bibinfo{year}{2015}\natexlab{}.
\newblock \showarticletitle{Heracles: Improving Resource Efficiency at Scale}.
  In \bibinfo{booktitle}{\emph{Proceedings of the 42nd Annual International
  Symposium on Computer Architecture (ISCA'15)}}.
\newblock


\bibitem[\protect\citeauthoryear{Mailthoday, Qureshi, Liang, Feng, de~Gonzalo,
  Li, Franke, Xiong, Huang, and Hwu}{Mailthoday et~al\mbox{.}}{2019}]%
        {deepstore:micro2019}
\bibfield{author}{\bibinfo{person}{Vikram~Sharma Mailthoday},
  \bibinfo{person}{Zaid Qureshi}, \bibinfo{person}{Weixin Liang},
  \bibinfo{person}{Ziyan Feng}, \bibinfo{person}{Simon~Garcia de Gonzalo},
  \bibinfo{person}{Youjie Li}, \bibinfo{person}{Hubertus Franke},
  \bibinfo{person}{Jinjun Xiong}, \bibinfo{person}{Jian Huang}, {and}
  \bibinfo{person}{Wenmei Hwu}.} \bibinfo{year}{2019}\natexlab{}.
\newblock \showarticletitle{{DeepStore: In-Storage Acceleration for Intelligent
  Queries}}. In \bibinfo{booktitle}{\emph{Proceedings of the 52nd IEEE/ACM
  International Symposium on Microarchitecture (MICRO'19)}}.
  \bibinfo{address}{Columbus, OH}.
\newblock


\bibitem[\protect\citeauthoryear{Matam, Koo, Zha, Tseng, and Annavaram}{Matam
  et~al\mbox{.}}{2019}]%
        {graphssd:isca2019}
\bibfield{author}{\bibinfo{person}{Kiran~Kumar Matam}, \bibinfo{person}{Gunjae
  Koo}, \bibinfo{person}{Haipeng Zha}, \bibinfo{person}{Hung-Wei Tseng}, {and}
  \bibinfo{person}{Murali Annavaram}.} \bibinfo{year}{2019}\natexlab{}.
\newblock \showarticletitle{{GraphSSD: Graph Semantics Aware SSD}}. In
  \bibinfo{booktitle}{\emph{Proceedings of the 46th Annual International
  Symposium on Computer Architecture (ISCA'19)}}. \bibinfo{address}{Phoenix,
  AZ}.
\newblock


\bibitem[\protect\citeauthoryear{{McCune}, {Li}, {Qu}, {Zhou}, {Datta},
  {Gligor}, and {Perrig}}{{McCune} et~al\mbox{.}}{2010}]%
        {trustvisor}
\bibfield{author}{\bibinfo{person}{J.~M. {McCune}}, \bibinfo{person}{Y. {Li}},
  \bibinfo{person}{N. {Qu}}, \bibinfo{person}{Z. {Zhou}}, \bibinfo{person}{A.
  {Datta}}, \bibinfo{person}{V. {Gligor}}, {and} \bibinfo{person}{A.
  {Perrig}}.} \bibinfo{year}{2010}\natexlab{}.
\newblock \showarticletitle{TrustVisor: Efficient TCB Reduction and
  Attestation}. In \bibinfo{booktitle}{\emph{2010 IEEE Symposium on Security
  and Privacy (Oakland'10)}}.
\newblock


\bibitem[\protect\citeauthoryear{McCune, Parno, Perrig, Reiter, and
  Isozaki}{McCune et~al\mbox{.}}{2008}]%
        {flicker}
\bibfield{author}{\bibinfo{person}{Jonathan~M. McCune},
  \bibinfo{person}{Bryan~J. Parno}, \bibinfo{person}{Adrian Perrig},
  \bibinfo{person}{Michael~K. Reiter}, {and} \bibinfo{person}{Hiroshi
  Isozaki}.} \bibinfo{year}{2008}\natexlab{}.
\newblock \showarticletitle{Flicker: An Execution Infrastructure for Tcb
  Minimization}. In \bibinfo{booktitle}{\emph{Proceedings of the 3rd ACM
  SIGOPS/EuroSys European Conference on Computer Systems (EuroSys'08)}}.
\newblock


\bibitem[\protect\citeauthoryear{Micron}{Micron}{2019}]%
        {microntlc}
\bibfield{author}{\bibinfo{person}{Micron}.} \bibinfo{year}{2019}\natexlab{}.
\newblock \bibinfo{title}{Micron 3D NAND Flash Memory}.
\newblock
\newblock


\bibitem[\protect\citeauthoryear{Muralimanohar, Balasubramonian, and
  Jouppi}{Muralimanohar et~al\mbox{.}}{2009}]%
        {cacti}
\bibfield{author}{\bibinfo{person}{Naveen Muralimanohar},
  \bibinfo{person}{Rajeev Balasubramonian}, {and} \bibinfo{person}{Norman~P
  Jouppi}.} \bibinfo{year}{2009}\natexlab{}.
\newblock \showarticletitle{{CACTI 6.0: A Tool to Model Large Caches}}.
\newblock \bibinfo{journal}{\emph{HP laboratories}} (\bibinfo{year}{2009}).
\newblock


\bibitem[\protect\citeauthoryear{Oleksenko, Trach, Krahn, Silberstein, and
  Fetzer}{Oleksenko et~al\mbox{.}}{2018}]%
        {varys:atc2018}
\bibfield{author}{\bibinfo{person}{Oleksii Oleksenko}, \bibinfo{person}{Bohdan
  Trach}, \bibinfo{person}{Robert Krahn}, \bibinfo{person}{Mark Silberstein},
  {and} \bibinfo{person}{Christof Fetzer}.} \bibinfo{year}{2018}\natexlab{}.
\newblock \showarticletitle{Varys: Protecting {SGX} Enclaves from Practical
  Side-Channel Attacks}. In \bibinfo{booktitle}{\emph{2018 {USENIX} Annual
  Technical Conference ({USENIX} {ATC}'18)}}. \bibinfo{address}{Boston, MA}.
\newblock


\bibitem[\protect\citeauthoryear{{Priebe}, {Vaswani}, and {Costa}}{{Priebe}
  et~al\mbox{.}}{2018}]%
        {enclavedb}
\bibfield{author}{\bibinfo{person}{C. {Priebe}}, \bibinfo{person}{K.
  {Vaswani}}, {and} \bibinfo{person}{M. {Costa}}.}
  \bibinfo{year}{2018}\natexlab{}.
\newblock \showarticletitle{EnclaveDB: A Secure Database Using SGX}. In
  \bibinfo{booktitle}{\emph{2018 IEEE Symposium on Security and Privacy
  (Oakland'18)}}.
\newblock


\bibitem[\protect\citeauthoryear{Reardon, Capkun, and Basin}{Reardon
  et~al\mbox{.}}{2012}]%
        {denfs}
\bibfield{author}{\bibinfo{person}{Joel Reardon}, \bibinfo{person}{Srdjan
  Capkun}, {and} \bibinfo{person}{David Basin}.}
  \bibinfo{year}{2012}\natexlab{}.
\newblock \showarticletitle{Data Node Encrypted File System: Efficient Secure
  Deletion for Flash Memory}. In \bibinfo{booktitle}{\emph{Presented as part of
  the 21st {USENIX} Security Symposium ({USENIX} Security'12)}}.
  \bibinfo{address}{Bellevue, WA}.
\newblock


\bibitem[\protect\citeauthoryear{Rogers, Chhabra, Prvulovic, and
  Solihin}{Rogers et~al\mbox{.}}{2007}]%
        {bmt:micro07}
\bibfield{author}{\bibinfo{person}{Brian Rogers}, \bibinfo{person}{Siddhartha
  Chhabra}, \bibinfo{person}{Milos Prvulovic}, {and} \bibinfo{person}{Yan
  Solihin}.} \bibinfo{year}{2007}\natexlab{}.
\newblock \showarticletitle{Using address independent seed encryption and
  bonsai merkle trees to make secure processors os-and performance-friendly}.
  In \bibinfo{booktitle}{\emph{40th Annual IEEE/ACM International Symposium on
  Microarchitecture (MICRO 2007)}}. IEEE, \bibinfo{pages}{183--196}.
\newblock


\bibitem[\protect\citeauthoryear{Ruan, He, and Cong}{Ruan
  et~al\mbox{.}}{2019}]%
        {insider:atc19}
\bibfield{author}{\bibinfo{person}{Zhenyuan Ruan}, \bibinfo{person}{Tong He},
  {and} \bibinfo{person}{Jason Cong}.} \bibinfo{year}{2019}\natexlab{}.
\newblock \showarticletitle{{INSIDER}: Designing In-Storage Computing System
  for Emerging High-Performance Drive}. In \bibinfo{booktitle}{\emph{2019
  {USENIX} Annual Technical Conference ({USENIX} {ATC} 19)}}.
  \bibinfo{publisher}{{USENIX} Association}, \bibinfo{address}{Renton, WA},
  \bibinfo{pages}{379--394}.
\newblock
\showISBNx{978-1-939133-03-8}
\urldef\tempurl%
\url{https://www.usenix.org/conference/atc19/presentation/ruan}
\showURL{%
\tempurl}


\bibitem[\protect\citeauthoryear{Saileshwar, Nair, Ramrakhyani, Elsasser, Joao,
  and Qureshi}{Saileshwar et~al\mbox{.}}{2018}]%
        {morphable:micro18}
\bibfield{author}{\bibinfo{person}{Gururaj Saileshwar},
  \bibinfo{person}{Prashant Nair}, \bibinfo{person}{Prakash Ramrakhyani},
  \bibinfo{person}{Wendy Elsasser}, \bibinfo{person}{Jose Joao}, {and}
  \bibinfo{person}{Moinuddin Qureshi}.} \bibinfo{year}{2018}\natexlab{}.
\newblock \showarticletitle{Morphable counters: Enabling compact integrity
  trees for low-overhead secure memories}. In \bibinfo{booktitle}{\emph{2018
  51st Annual IEEE/ACM International Symposium on Microarchitecture (MICRO)}}.
  IEEE, \bibinfo{pages}{416--427}.
\newblock


\bibitem[\protect\citeauthoryear{Samsung}{Samsung}{2020}]%
        {www:smartssd}
\bibfield{author}{\bibinfo{person}{Samsung}.} \bibinfo{year}{2020}\natexlab{}.
\newblock \bibinfo{title}{SmartSSD Computational Storage Drive}.
\newblock
  \bibinfo{howpublished}{\\\url{https://samsungsemiconductor-us.com/smartssd/index.html}}.
\newblock


\bibitem[\protect\citeauthoryear{Santos, Raj, Saroiu, and Wolman}{Santos
  et~al\mbox{.}}{2014}]%
        {trustzone:asplos14}
\bibfield{author}{\bibinfo{person}{Nuno Santos}, \bibinfo{person}{Himanshu
  Raj}, \bibinfo{person}{Stefan Saroiu}, {and} \bibinfo{person}{Alec Wolman}.}
  \bibinfo{year}{2014}\natexlab{}.
\newblock \showarticletitle{Using ARM Trustzone to Build a Trusted Language
  Runtime for Mobile Applications}. In \bibinfo{booktitle}{\emph{Proceedings of
  the 19th International Conference on Architectural Support for Programming
  Languages and Operating Systems (ASPLOS'14)}}.
\newblock


\bibitem[\protect\citeauthoryear{{Schuster}, {Costa}, {Fournet}, {Gkantsidis},
  {Peinado}, {Mainar-Ruiz}, and {Russinovich}}{{Schuster}
  et~al\mbox{.}}{2015}]%
        {vc3:oakland2015}
\bibfield{author}{\bibinfo{person}{F. {Schuster}}, \bibinfo{person}{M.
  {Costa}}, \bibinfo{person}{C. {Fournet}}, \bibinfo{person}{C. {Gkantsidis}},
  \bibinfo{person}{M. {Peinado}}, \bibinfo{person}{G. {Mainar-Ruiz}}, {and}
  \bibinfo{person}{M. {Russinovich}}.} \bibinfo{year}{2015}\natexlab{}.
\newblock \showarticletitle{VC3: Trustworthy Data Analytics in the Cloud Using
  SGX}. In \bibinfo{booktitle}{\emph{2015 IEEE Symposium on Security and
  Privacy (Oakland'15)}}.
\newblock


\bibitem[\protect\citeauthoryear{{Security Flaws Found in Intel Software, Data
  Center SSDs}}{{Security Flaws Found in Intel Software, Data Center SSDs}}{[n.
  d.]}]%
        {ssdflaw}
\bibfield{author}{\bibinfo{person}{{Security Flaws Found in Intel Software,
  Data Center SSDs}}.} \bibinfo{year}{[n. d.]}\natexlab{}.
\newblock
  \bibinfo{howpublished}{\url{https://www.tomshardware.com/news/intel-security-vulnerabilities-processor-diagnostic-tool-ssd,39845.html}}.
\newblock


\bibitem[\protect\citeauthoryear{Seshadri, Gahagan, Bhaskaran, Bunker, De, Jin,
  Liu, and Swanson}{Seshadri et~al\mbox{.}}{2014}]%
        {willow}
\bibfield{author}{\bibinfo{person}{Sudharsan Seshadri}, \bibinfo{person}{Mark
  Gahagan}, \bibinfo{person}{Sundaram Bhaskaran}, \bibinfo{person}{Trevor
  Bunker}, \bibinfo{person}{Arup De}, \bibinfo{person}{Yanqin Jin},
  \bibinfo{person}{Yang Liu}, {and} \bibinfo{person}{Steven Swanson}.}
  \bibinfo{year}{2014}\natexlab{}.
\newblock \showarticletitle{{Willow: A User-programmable SSD}}. In
  \bibinfo{booktitle}{\emph{Proceedings of the 11th USENIX Conference on
  Operating Systems Design and Implementation (OSDI'14)}}.
  \bibinfo{address}{Broomfield, CO}.
\newblock


\bibitem[\protect\citeauthoryear{Shilov}{Shilov}{2019}]%
        {riscv-samsung}
\bibfield{author}{\bibinfo{person}{Anton Shilov}.}
  \bibinfo{year}{2019}\natexlab{}.
\newblock \bibinfo{title}{Samsung to Use SiFive RISC-V Cores for SoCs,
  Automotive, 5G Applications}.
\newblock
  \bibinfo{howpublished}{\\\url{https://www.anandtech.com/show/15228/samsung-to-use-riscv-cores}}.
\newblock


\bibitem[\protect\citeauthoryear{Shue and Freedman}{Shue and Freedman}{2014}]%
        {libra:eurosys2014}
\bibfield{author}{\bibinfo{person}{David Shue} {and}
  \bibinfo{person}{Michael~J. Freedman}.} \bibinfo{year}{2014}\natexlab{}.
\newblock \showarticletitle{From Application Requests to Virtual IOPs:
  Provisioned Key-Value Storage with Libra}. In
  \bibinfo{booktitle}{\emph{Proceedings of the Ninth European Conference on
  Computer Systems (EuroSys'14)}}. \bibinfo{address}{New York, NY, USA}.
\newblock


\bibitem[\protect\citeauthoryear{Shue, Freedman, and Shaikh}{Shue
  et~al\mbox{.}}{2012}]%
        {cloudstorage:osdi2012}
\bibfield{author}{\bibinfo{person}{David Shue}, \bibinfo{person}{Michael~J.
  Freedman}, {and} \bibinfo{person}{Anees Shaikh}.}
  \bibinfo{year}{2012}\natexlab{}.
\newblock \showarticletitle{Performance Isolation and Fairness for Multi-Tenant
  Cloud Storage}. In \bibinfo{booktitle}{\emph{Proceedings of the 10th {USENIX}
  Symposium on Operating Systems Design and Implementation ({OSDI} 12)}}.
  \bibinfo{address}{Hollywood, CA}.
\newblock


\bibitem[\protect\citeauthoryear{Shukla, Thota, Raman, Gajendran, Shah, Ziuzin,
  Sundaram, Guajardo, Wawrzyniak, Boshra, Ferreira, Nassar, Koltachev, Huang,
  Sengupta, Levandoski, and Lomet}{Shukla et~al\mbox{.}}{2015}]%
        {documentdb:vldb2015}
\bibfield{author}{\bibinfo{person}{Dharma Shukla}, \bibinfo{person}{Shireesh
  Thota}, \bibinfo{person}{Karthik Raman}, \bibinfo{person}{Madhan Gajendran},
  \bibinfo{person}{Ankur Shah}, \bibinfo{person}{Sergii Ziuzin},
  \bibinfo{person}{Krishnan Sundaram}, \bibinfo{person}{Miguel~Gonzalez
  Guajardo}, \bibinfo{person}{Anna Wawrzyniak}, \bibinfo{person}{Samer Boshra},
  \bibinfo{person}{Renato Ferreira}, \bibinfo{person}{Mohamed Nassar},
  \bibinfo{person}{Michael Koltachev}, \bibinfo{person}{Ji Huang},
  \bibinfo{person}{Sudipta Sengupta}, \bibinfo{person}{Justin Levandoski},
  {and} \bibinfo{person}{David Lomet}.} \bibinfo{year}{2015}\natexlab{}.
\newblock \showarticletitle{Schema-Agnostic Indexing with Azure DocumentDB}.
\newblock \bibinfo{journal}{\emph{Proceeding of VLDB Endow.}}
  \bibinfo{volume}{8}, \bibinfo{number}{12} (\bibinfo{date}{Aug.}
  \bibinfo{year}{2015}).
\newblock


\bibitem[\protect\citeauthoryear{Sirer, de~Bruijn, Reynolds, Shieh, Walsh,
  Williams, and Schneider}{Sirer et~al\mbox{.}}{2011}]%
        {logicalattestation}
\bibfield{author}{\bibinfo{person}{Emin~G\"{u}n Sirer}, \bibinfo{person}{Willem
  de Bruijn}, \bibinfo{person}{Patrick Reynolds}, \bibinfo{person}{Alan Shieh},
  \bibinfo{person}{Kevin Walsh}, \bibinfo{person}{Dan Williams}, {and}
  \bibinfo{person}{Fred~B. Schneider}.} \bibinfo{year}{2011}\natexlab{}.
\newblock \showarticletitle{Logical Attestation: An Authorization Architecture
  for Trustworthy Computing}. In \bibinfo{booktitle}{\emph{Proceedings of the
  Twenty-Third ACM Symposium on Operating Systems Principles (SOSP'11)}}.
\newblock


\bibitem[\protect\citeauthoryear{{Stavros Volos and Kapil Vaswani and Rodrigo
  Bruno}}{{Stavros Volos and Kapil Vaswani and Rodrigo Bruno}}{2018}]%
        {graviton:osdi2018}
\bibfield{author}{\bibinfo{person}{{Stavros Volos and Kapil Vaswani and Rodrigo
  Bruno}}.} \bibinfo{year}{2018}\natexlab{}.
\newblock \showarticletitle{{Graviton: Trusted Execution Environments on
  GPUs}}. In \bibinfo{booktitle}{\emph{Proceedings of the 13th USENIX Symposium
  on Operating Systems Design and Implementation (OSDI'18)}}.
  \bibinfo{address}{Carlsbad, CA}.
\newblock


\bibitem[\protect\citeauthoryear{{Szekeres}, {Payer}, {Wei}, and
  {Song}}{{Szekeres} et~al\mbox{.}}{2013}]%
        {sok}
\bibfield{author}{\bibinfo{person}{L. {Szekeres}}, \bibinfo{person}{M.
  {Payer}}, \bibinfo{person}{T. {Wei}}, {and} \bibinfo{person}{D. {Song}}.}
  \bibinfo{year}{2013}\natexlab{}.
\newblock \showarticletitle{SoK: Eternal War in Memory}. In
  \bibinfo{booktitle}{\emph{Proceedings of the 2013 IEEE Symposium on Security
  and Privacy (Oakland'13)}}.
\newblock


\bibitem[\protect\citeauthoryear{Taassori, Shafiee, and
  Balasubramonian}{Taassori et~al\mbox{.}}{2018}]%
        {vault:asplos18}
\bibfield{author}{\bibinfo{person}{Meysam Taassori}, \bibinfo{person}{Ali
  Shafiee}, {and} \bibinfo{person}{Rajeev Balasubramonian}.}
  \bibinfo{year}{2018}\natexlab{}.
\newblock \showarticletitle{VAULT: Reducing paging overheads in SGX with
  efficient integrity verification structures}. In
  \bibinfo{booktitle}{\emph{Proceedings of the Twenty-Third International
  Conference on Architectural Support for Programming Languages and Operating
  Systems}}. \bibinfo{pages}{665--678}.
\newblock


\bibitem[\protect\citeauthoryear{Tiwari, Boboila, Vazhkudai, Kim, Ma,
  Desnoyers, and Solihin}{Tiwari et~al\mbox{.}}{2013}]%
        {Tiwari:2013:AFT:2591272.2591286}
\bibfield{author}{\bibinfo{person}{Devesh Tiwari}, \bibinfo{person}{Simona
  Boboila}, \bibinfo{person}{Sudharshan~S. Vazhkudai},
  \bibinfo{person}{Youngjae Kim}, \bibinfo{person}{Xiaosong Ma},
  \bibinfo{person}{Peter~J. Desnoyers}, {and} \bibinfo{person}{Yan Solihin}.}
  \bibinfo{year}{2013}\natexlab{}.
\newblock \showarticletitle{Active Flash: Towards Energy-efficient, In-situ
  Data Analytics on Extreme-scale Machines}. In
  \bibinfo{booktitle}{\emph{Proceedings of the 11th USENIX Conference on File
  and Storage Technologies (FAST'13)}}. \bibinfo{address}{San Jose, CA}.
\newblock


\bibitem[\protect\citeauthoryear{Tiwari, Vazhkudai, Kim, Ma, Boboila, and
  Desnoyers}{Tiwari et~al\mbox{.}}{2012}]%
        {Tiwari:2012:RDM:2387869.2387873}
\bibfield{author}{\bibinfo{person}{Devesh Tiwari},
  \bibinfo{person}{Sudharshan~S. Vazhkudai}, \bibinfo{person}{Youngjae Kim},
  \bibinfo{person}{Xiaosong Ma}, \bibinfo{person}{Simona Boboila}, {and}
  \bibinfo{person}{Peter~J. Desnoyers}.} \bibinfo{year}{2012}\natexlab{}.
\newblock \showarticletitle{{Reducing Data Movement Costs Using Energy
  Efficient, Active Computation on SSD}}. In
  \bibinfo{booktitle}{\emph{Proceedings of the 2012 USENIX Conference on
  Power-Aware Computing and Systems (HotPower'12)}}.
  \bibinfo{address}{Hollywood, CA}.
\newblock


\bibitem[\protect\citeauthoryear{Tseng, Grupp, and Swanson}{Tseng
  et~al\mbox{.}}{2013}]%
        {tseng:dac2013}
\bibfield{author}{\bibinfo{person}{Hung-Wei Tseng}, \bibinfo{person}{Laura~M.
  Grupp}, {and} \bibinfo{person}{Steven Swanson}.}
  \bibinfo{year}{2013}\natexlab{}.
\newblock \showarticletitle{Underpowering NAND Flash: Profits and Perils}. In
  \bibinfo{booktitle}{\emph{Proceedings of the 50th Annual Design Automation
  Conference (DAC'13)}}.
\newblock


\bibitem[\protect\citeauthoryear{University}{University}{2020}]%
        {openssd-web}
\bibfield{author}{\bibinfo{person}{Hanyang University}.}
  \bibinfo{year}{2020}\natexlab{}.
\newblock \bibinfo{title}{The OpenSSD Project: Open-Source Solid-State Drive
  Project for Research and Education}.
\newblock
\newblock


\bibitem[\protect\citeauthoryear{Wang, Zhou, Coats, and Huang}{Wang
  et~al\mbox{.}}{2019}]%
        {timessd:eurosys2019}
\bibfield{author}{\bibinfo{person}{Xiaohao Wang}, \bibinfo{person}{You Zhou},
  \bibinfo{person}{Chance~C. Coats}, {and} \bibinfo{person}{Jian Huang}.}
  \bibinfo{year}{2019}\natexlab{}.
\newblock \showarticletitle{{Project Almanac: A Time-Traveling Solid-State
  Drive}}. In \bibinfo{booktitle}{\emph{Proceedings of the 14th European
  Conference on Computer Systems (EuroSys'19)}}. \bibinfo{address}{Dresden,
  Germany}.
\newblock


\bibitem[\protect\citeauthoryear{Weis}{Weis}{2014}]%
        {physicalattack:blackhat2014}
\bibfield{author}{\bibinfo{person}{Steve Weis}.}
  \bibinfo{year}{2014}\natexlab{}.
\newblock \showarticletitle{{Protecting Data In-Use from Firmware and Physical
  Attacks}}.
\newblock \bibinfo{journal}{\emph{Proceedings of Black Hat}}
  (\bibinfo{year}{2014}).
\newblock


\bibitem[\protect\citeauthoryear{{Weng}, {Liu}, {Dadu}, {Wang}, {Shah}, and
  {Nowatzki}}{{Weng} et~al\mbox{.}}{2020}]%
        {dsagen:isca2020}
\bibfield{author}{\bibinfo{person}{J. {Weng}}, \bibinfo{person}{S. {Liu}},
  \bibinfo{person}{V. {Dadu}}, \bibinfo{person}{Z. {Wang}}, \bibinfo{person}{P.
  {Shah}}, {and} \bibinfo{person}{T. {Nowatzki}}.}
  \bibinfo{year}{2020}\natexlab{}.
\newblock \showarticletitle{DSAGEN: Synthesizing Programmable Spatial
  Accelerators}. In \bibinfo{booktitle}{\emph{Proceedings of the ACM/IEEE 47th
  Annual International Symposium on Computer Architecture (ISCA'20)}}.
\newblock


\bibitem[\protect\citeauthoryear{{What is a buffer overflow? And how hackers
  exploit these vulnerabilities}}{{What is a buffer overflow? And how hackers
  exploit these vulnerabilities}}{[n. d.]}]%
        {bufferoverflow}
\bibfield{author}{\bibinfo{person}{{What is a buffer overflow? And how hackers
  exploit these vulnerabilities}}.} \bibinfo{year}{[n. d.]}\natexlab{}.
\newblock
  \bibinfo{howpublished}{\url{https://www.csoonline.com/article/3513477/what-is-a-buffer-overflow-and-how-hackers-exploit-these-vulnerabilities.html}}.
\newblock


\bibitem[\protect\citeauthoryear{Wile}{Wile}{2014}]%
        {capi:2014:whitepaper}
\bibfield{author}{\bibinfo{person}{Bruce Wile}.}
  \bibinfo{year}{2014}\natexlab{}.
\newblock \showarticletitle{{Coherent Accelerator Processor Interface (CAPI)
  for POWER8 Systems}}.
\newblock \bibinfo{journal}{\emph{White Paper}} (\bibinfo{date}{Sep}
  \bibinfo{year}{2014}).
\newblock


\bibitem[\protect\citeauthoryear{Yan, Englender, Prvulovic, Rogers, and
  Solihin}{Yan et~al\mbox{.}}{2006}]%
        {encryption:isca06}
\bibfield{author}{\bibinfo{person}{Chenyu Yan}, \bibinfo{person}{Daniel
  Englender}, \bibinfo{person}{Milos Prvulovic}, \bibinfo{person}{Brian
  Rogers}, {and} \bibinfo{person}{Yan Solihin}.}
  \bibinfo{year}{2006}\natexlab{}.
\newblock \showarticletitle{Improving cost, performance, and security of memory
  encryption and authentication}.
\newblock \bibinfo{journal}{\emph{ACM SIGARCH Computer Architecture News}}
  \bibinfo{volume}{34}, \bibinfo{number}{2} (\bibinfo{year}{2006}),
  \bibinfo{pages}{179--190}.
\newblock


\bibitem[\protect\citeauthoryear{Zhang, Xiong, Xu, Shu, Li, Cheng, Chen, and
  Moscibroda}{Zhang et~al\mbox{.}}{2017}]%
        {fpgacloud}
\bibfield{author}{\bibinfo{person}{Jiansong Zhang}, \bibinfo{person}{Yongqiang
  Xiong}, \bibinfo{person}{Ningyi Xu}, \bibinfo{person}{Ran Shu},
  \bibinfo{person}{Bojie Li}, \bibinfo{person}{Peng Cheng},
  \bibinfo{person}{Guo Chen}, {and} \bibinfo{person}{Thomas Moscibroda}.}
  \bibinfo{year}{2017}\natexlab{}.
\newblock \showarticletitle{The Feniks FPGA Operating System for Cloud
  Computing}. In \bibinfo{booktitle}{\emph{Proceedings of the 8th Asia-Pacific
  Workshop on Systems (APSys'17)}}.
\newblock


\bibitem[\protect\citeauthoryear{Zhang, Tatemura, Patel, and Hacigumus}{Zhang
  et~al\mbox{.}}{2014}]%
        {multissd:sigmod2014}
\bibfield{author}{\bibinfo{person}{Ning Zhang}, \bibinfo{person}{Junichi
  Tatemura}, \bibinfo{person}{Jignesh Patel}, {and} \bibinfo{person}{Hakan
  Hacigumus}.} \bibinfo{year}{2014}\natexlab{}.
\newblock \showarticletitle{Re-Evaluating Designs for Multi-Tenant OLTP
  Workloads on SSD-BasedI/O Subsystems}. In
  \bibinfo{booktitle}{\emph{Proceedings of the 2014 ACM SIGMOD International
  Conference on Management of Data (SIGMOD'14)}}.
\newblock


\bibitem[\protect\citeauthoryear{Zheng, Dave, Beekman, Popa, Gonzalez, and
  Stoica}{Zheng et~al\mbox{.}}{2017}]%
        {opaque:nsdi17}
\bibfield{author}{\bibinfo{person}{Wenting Zheng}, \bibinfo{person}{Ankur
  Dave}, \bibinfo{person}{Jethro~G. Beekman}, \bibinfo{person}{Raluca~Ada
  Popa}, \bibinfo{person}{Joseph~E. Gonzalez}, {and} \bibinfo{person}{Ion
  Stoica}.} \bibinfo{year}{2017}\natexlab{}.
\newblock \showarticletitle{Opaque: An Oblivious and Encrypted Distributed
  Analytics Platform}. In \bibinfo{booktitle}{\emph{14th {USENIX} Symposium on
  Networked Systems Design and Implementation ({NSDI}'17)}}.
  \bibinfo{address}{Boston, MA}.
\newblock


\end{thebibliography}
